\documentclass[useAMS,usenatbib]{mn2e} 
\usepackage{graphicx}
\usepackage{subfigure}
\usepackage{epsfig}
\usepackage{times}
\usepackage{natbib}
\usepackage{amsfonts}
\usepackage{amsmath}
\usepackage{amsbsy}

\newcommand{\bhl}{}
\newcommand{\ehl}{}
\newcommand{\bhm}{}
\newcommand{\ehm}{}

\newcommand{\beq}{\begin{equation}}
\newcommand{\eeq}{\end{equation}}
\newcommand{\rd}{{\rm d}}

\newif\ifAMStwofonts

\def\rin{$r_{\rm in}$}
\def\rout{$r_{\rm out}$}
\def\rc{$r_{\rm c}$}
\def\del2r{$\Delta r_{\rm 2}$}
\def\delr{$\Delta r$}
\def\ddr{\frac{\partial}{\partial r}}

\title{Long-term evolution of discs around magnetic stars}
\author[C.\ R.\ D'Angelo \& H.\ C.\ Spruit]
       {Caroline R.\ D'Angelo and
         Hendrik C.\ Spruit \\  Max Planck Institute for
         Astrophysics, Karl-Schwarzchildstr.1 D-85741 Garching, Germany} 
\date{\today}

\pagerange{\pageref{firstpage}--\pageref{lastpage}}
\pubyear{2011}

\begin{document}

\maketitle
\label{firstpage}
\begin{abstract}
  We investigate the evolution of a thin viscous disc surrounding
  magnetic star, including the spindown of the star by the magnetic
  torques it exerts on the disc. The transition from an accreting to a
  non-accreting state, and the change of the magnetic torque across
  the corotation radius \rc\ are included in a generic way, the widths
  of the transition taken in the range suggested by numerical
  simulations. In addition to the standard accreting state, two more
  are found. An accreting state can develop into a `dead' disc state,
  with inner edge \rin\ well outside corotation. More often, a
  `trapped' state develops, in which \rin\ stays close to corotation
  even at very low accretion rates. The long-term evolution of these
  two states is different. In the dead state the star spins down
  incompletely, retaining much of its initial spin. In the trapped
  state the star asymptotically can spin down to arbitarily low rates,
  its angular momentum transferred to the disc. We identify these
  outcomes with respectively the rapidly rotating and the very slowly
  rotating classes of Ap stars and magnetic white dwarfs.
\end{abstract}

\section{Introduction}
Accreting stars with strong magnetic fields are generally observed to
rotate more slowly than their less-magnetic or discless
counterparts. In protostars, T Tauri systems (which often have strong
surface magnetic fields of $\sim 10^2-10^3$ G) with discs rotate more
rapidly than systems without discs \citep{2008ApJ...688..437G}. Most
(but not all) mainsequence Ap stars (with surface fields of up to
$10^4$ G) are observed to rotate much slower than normal A stars
\citep{2002A&A...384..554S}, and recent work has suggests this
relationship extends down to their pre-main-sequence progenitors, the
Herbig Ae stars \citep{2008CoSka..38..235A}. In high energy systems
the result is similar: accreting neutron stars with weak ($\sim10^8$
G) fields rotate up to $10^4$ times faster than neutron stars with
strong ($\sim 10^{12}$ G) fields. These observations suggest that the
interaction between the accretion disc and stellar magnetic field
plays a critical role in regulating the spin-rate of the star.

Early theoretical studies of accretion predicted that a strong stellar
field would truncate the accretion disc some distance from the surface
of the star, with the truncation radius located roughly where the
magnetic pressure ($B^2/4\pi$) equals the ram pressure of the
infalling gas ($\dot{m}v_{\rm r}/2\pi r^2$), so that infalling matter
is channelled onto the surface via magnetic field lines, causing the
star to spin up.  \citep{1972A&A....21....1P}. This assumes that the
disc is truncated inside the corotation radius ($r_{\rm c} \equiv
\left(GM_*/\Omega^2_*\right)^{1/3}$), where the star's spin
frequency is equal to the disc's Keplerian frequency). If instead the
magnetic field spins faster than the inner edge of the disc, 
a centrifugal barrier \bhl prevents accretion\ehl.
Interaction between the magnetic field and the disc will then spin 
down the star \citep{1975A&A....39..185I,1991MNRAS.251..555M,
1999ApJ...514..368L, 2004ApJ...616L.151R,2006ApJ...646..304U}.

\bhl The presence of the centrifugal barrier is often equated in the literature with 
the idea that the accreting gas will be flung out, or  `propellered' out of the system 
so as to maintain a steady state. This assumption turns out to be 
both arbitrary and unnecessary. For example, in order for the accreting 
material to be flung out of the system, the disc must be truncated a sufficient 
distance away from \rc. Otherwise the rotational velocity difference between 
the disc and the magnetosphere is too small (\citealt{1993ApJ...402..593S}).

Steady disc solutions with a centrifugal barrier at the inner edge were 
first described by \cite{1977PAZh....3..262S}, who called them `dead discs', 
because even though the disc is actively transporting angular momentum
outwards, no accretion onto the star takes place and the disc itself
is very dim. 

As pointed out already in \cite{1977PAZh....3..262S} and 
\cite{1993ApJ...402..593S}, in a system with an externally imposed mass 
flux the likely effect of a centrifugal barrier is to cause the accretion onto 
the star to be {\em cyclic}. Accretion phases alternate with quiescent periods 
during which mass piles up outside the barrier, without mass having to leave 
the system. In the quiescent phase\ehl, the angular momentum 
extracted from the star by the disc-field interaction is carried outward 
through the disc by viscous stress. This alters the surface density 
profile of the disc from the usual accreting solution.



\bhl In our previous paper (\citealt{2010MNRAS.406.1208D}; hereafter DS10)
we studied this form of cyclic accretion with numerical solutions of
the viscous diffusion equation for a thin disc subject to a magnetic
torque. As in the (somewhat more ad hoc) model of \cite{1993ApJ...402..593S}, limit cycles
of the relaxation oscillator type were found\ehl.  The cycle period of
these oscillations depends on the accretion rate, from fast
oscillations at higher mass flux to arbitrarily long periods at low
accretion rates.



Instead of the two states: accreting and dead as suggested above, 
the results in DS10 are actually described better by including a third,
intermediate state we call here the `trapped' state:
\begin{enumerate}
\item{\rin$<$\rc: accreting state, star spins up,}
\item{\rc$-\Delta<$\rin$<$\rc$+\Delta$: trapped state, spinup or spindown,}
\item{\rin-\rc$\gg\Delta$ star spins down, no accretion (dead disc)},
\end{enumerate}
where $\Delta\ll$\rc\ is a narrow range around corotation, to be specified
later. In state (ii), the inner edge of the disc remains close to corotation
over a range of accretion rates onto the star, and the net torque on the
star can be of either sign, depending on the precise location of the 
inner edge of the disc.

For a given accretion rate, a disc that starts in state (i) will gradually 
move into state (ii) or (iii) as the star spins up and $r_{\rm c}$ moves 
inward. In state (ii) accretion can proceed steadily or happen in bursts, 
depending on the disc-field interaction at \rin. \bhl For steady externally 
imposed accretion a disc in this state\ehl\ will eventually move into spin equilibrium 
with the star, so that the net torque on the star is zero. \bhl In the dead 
state (iii) a steady state can exist if the torque exterted by the star is taken 
up at the outer edge of the disc by a companion star\ehl.
If we neglect the transition to the propeller regime, then in
theory the dead disc solution can exist for a disc truncated at any
distance outside \rc. Such a disc will remain static as the star spins
down and \rc\ moves outward. Our model is thus qualitatively different
from the conventional `propeller' picture since at very low accretion 
rates a considerable amount of mass remains confined in the
disc, and the star can be efficiently spun down.



\bhl In the following we study the long-term evolution of the
star-disc system by using the description of magnetospheric accretion
in DS10, allowing the star's spin rate to evolve. Of special interest
will be the trapped state (ii), since in many cases the evolution of
the system ends in it. The accretion cycles found in DS10 also take
place essentially within a trapped state. The inner edge of the disc
is near corotation in the trapped state, as is the case also for a
disc in spin equilibrium with the accreting star. Spin equilibrium is
only a special case of a trapped state, however. In general a trapped
state is not one of spin equilibrium, spinup is possible as well as
spindown. We focus particular attention on cases in which the mean
accretion rate through the disc is very low and the star cannot easily
move into spin equilibrium with the disc. This is most applicable for
isolated stars, or accreting binaries in quiescence.

This scenario poses a number of questions which we address in the
course of this paper. These include: under what conditions does the
disc get into a trapped state, and when does it instead evolve into a
dead state? It will turn out that this is\ehl\
determined by the details of the disc-field interaction and the ratio
of the spin-down timescale of the star ($T_{\rm SD}$) to the viscous
timescale of the disc ($T_{\rm visc}$). The initial conditions of the
disc also significantly influence the outcome. In section
\ref{sec3:const_am} we ask how an initially trapped disc could become
untrapped as a dead disc state. In particular, does this depend on the
initial location of the inner edge of the disc, the initial accretion
rate, the presence or absence of a companion or the size of the disc?
Finally, in sec. \ref{sec3:trapuntrap} we discuss the physics that determines
whether a disc will become trapped, and \ref{sec3:Apstars} we ask
whether a trapped disc could plausibly regulate the slow spins
observed in Ap stars, some of which have spin periods of up to
decades.

In a companion paper we investigate the observable consequences of a
trapped disc, focusing in particular on how the burst instability
studied in DS10 will change the spin evolution and observable
properties of the star. In that paper we also discuss our model's
predictions in terms of observations of magnetospherically regulated
accretion in both protostars and X-ray binaries.

We use the code developed in DS10, adding the star's moment of inertia
as a parameter of the problem in order to follow the spin evolution of
the star in response to the disc interaction. We can then
simultaneously follow the viscous evolution of the disc and spin
evolution of the star as the star's spin changes, and explore how
these two interact with each other. We describe our model in more
detail in the following section.

\section{Magnetospheric interactions with a thin disc}
\subsection{Magnetic torque}
\label{sec3:model}
The interaction between a strong stellar magnetic field and
surrounding accretion disc truncates the disc close to the star, and
forces incoming matter to accrete along closed field lines onto the
surface of the star in a region called the magnetosphere. At the
outer edge of the magnetosphere (termed here the magnetospheric
radius), the field lines become strongly embedded in the disc over
some small radial extent that we term the {\em interaction region},
\delr. The differential rotation between the star and the Keplerian
disc will cause the field lines to be twisted, which will generate a
toroidal component to an initially poloidal field
(e.g. \citealt{1977ApJ...217..578G}). This will allow the transfer of
angular momentum between the disc and star, with the torque per unit
area exerted by the field on the disc given by ${\boldsymbol
  \tau} = rS_{z\phi}{\bf \hat{z}}$, where:
\begin{equation}
\label{eq3:stress}
S_{z\phi} \equiv  \frac{B_\phi B_z}{4\pi}
\end{equation}
is the magnetic stress generated by the twisted field lines. Both
theoretical arguments
(e.g. \citealt{1985A&A...143...19A,1995MNRAS.275..244L}) and numerical
simulations (such as \citealt{1997ApJ...489..890M,
  1997ApJ...489..199G, 1996ApJ...468L..37H}) suggest that the strong
coupling between magnetic field lines and the disc will cause the
field lines to inflate and open. The inflation and opening of field
lines limits the growth of the $B_\phi$ component for the field to
$B_\phi=\eta B_z$, with $\eta$ of order unity, and reduces the radial 
extent of the interaction region,
since beyond a given radius the field lines are always open and the
disc-field connection will be severed. We take the interaction region 
to be narrow, $\Delta r/r < 1$ (as found in numerical simulations,
see section 2 of  DS10 for a more detailed discussion).
\bhl Assuming the star's dipole field strength $B_{\rm d}(r)$ as an estimate of 
$B_z$, and  taking into account that $S$ acts on both sides of the disc, 
(\ref{eq3:stress}) yields the magnetic torque $T_0$ exterted on the disc:
\beq T_0=4\pi r\Delta r \,rS_{z\phi} = \eta r^2\Delta r B_{\rm d}^2.\label{TB} \eeq

This torque exists only if the inner edge \rin\ of the disc is outside the 
corotation radius \rc. For \rin\ $<$ \rc, we have instead a disc accreting 
on an object rotating slower than the Kepler rate at \rin. By the standard 
theory of thin viscous discs, the torque exterted on the disc by the accreting 
object then vanishes, independent of the nature of the object. The torque 
$T_B(r_{\rm i})$ thus changes over a narrow range around \rc. To model this 
transition we introduce a `connecting function' $y_\Sigma$:
\beq T_B(r_{\rm in} )=y_\Sigma(r_{\rm in}) T_0(r_{\rm in}), \label{BT}\eeq
with the properties $y_\Sigma\rightarrow 0$ ($r_{\rm c}-r_{\rm in}\gg\Delta r$), 
 $y_\Sigma\rightarrow 1$ ($r_{\rm in}-r_{\rm c} \gg\Delta r$). As in DS10, we 
 take for this function
\begin{equation}
\label{ys}
y_\Sigma = \frac{1}{2}\left[1 + \tanh\left(\frac{r_{\rm in}-r_{\rm c}}{\Delta r}\right)\right].
\end{equation}
The width of the transition is thus described by $\Delta r$. We take 
the same value for it as used in eq.\ (\ref{TB}).\ehl

\subsection{Model for disc-magnetosphere interaction}
\label{interact}
In DS10 we derived a description of the interaction between a disc and
magnetic-field for a disc truncated either inside or outside \rc, and
introduced two numerical parameters to connect the two regimes. To
keep the problem axisymmetric, we assumed a dipolar magnetic field,
with the dipole axis aligned with the stellar and disc rotation
axis. Since the region of interaction between the disc and the field
is small, we use our description of the interaction as a boundary
condition for a standard thin accretion disc
\citep{1973A&A....24..337S}.

To evolve a thin disc in time, we must choose a description for the
effective viscosity $(\nu)$ that allows transport of angular
momentum. We adopt an $\alpha$ prescription for the viscosity and
assume a constant scale height ($h$) for the disc, so that:
\begin{equation}
\label{eq3:visc}
\nu = \alpha (GM_*)^{1/2}(h/r)^2r^{1/2}.
\end{equation}
At the inner edge of the disc the behaviour is regulated by the
disc-field interaction. However, since the interaction region is
small, we incorporate the interaction as a boundary condition on the
inner disc, and assume that the majority of the disc is shielded from
the magnetic field and then evolves as a standard viscous disc, albeit
with a very different inner boundary condition from the standard
one. Below we summarize our analysis of the disc-field interaction and
how these translate into boundary conditions on the disc. (For the
detailed derivation of our boundary conditions,  see sections
2.3, 2.4, and 3.2 of DS10).

\subsubsection{Surface density at ${r_{\rm in}}$}
\label{sec3:sigma_rin}
In a dead disc, the disc-field interaction prevents matter from
accreting or being expelled from the system, instead retaining matter
that interacts with the magnetic field. This implies that the angular
momentum injected via magnetic torques in the interaction region
$\Delta r$ must be transported outwards by viscous torques in the
disc. A dead disc will therefore have a maximum in surface density at
\rin, and $\Sigma(r_{\rm in})$ will depend on the amount of angular
momentum being added by the disc-field interaction.

By equating the amount of angular momentum added by the field to the
amount carried outwards by viscous processes, we can calculate the
surface density at the inner boundary of the disc needed for the
injected angular momentum to be carried away. \bhl This yields (see
DS10) a value for the surface density $\Sigma$ at the inner edge of a
dead disc, proportional to the magnetic torque (\ref{BT})\ehl:
\begin{equation}
\label{eq3:sigma_rin}
3\pi\nu\Sigma(r_{\rm in}) = {T_B\over  r^2\Omega_{\rm K}}\bigg|_{r_{\rm in}},
\end{equation}
where $\Omega_{\rm K}$ is the Keplerian rotation frequency. 
If the stellar
field is a dipole and we use (\ref{eq3:visc}) to describe the
viscosity, then for $r_{\rm in} > r_{\rm c}$, $\Sigma(r_{\rm in})
\propto r^{-4}_{\rm in}$. \bhl$\Sigma(r_{\rm in})$ thus decreases rapidly 
with increasing $r_{\rm in}$.\ehl

  

\subsubsection{Accretion rate across $r_{\rm c}$}
\label{sec3:dtrin}

\bhl If the inner edge is well inside the corotation radius \rc,\  we use a standard 
result to estimate the location of \rin\ as a function of the accretion rate 
$\dot m_{\rm a}$. It is obtained from \ehl\ the azimuthal equation
of motion for gas at the point at which it is forced to corotate with
the star (c.f. \citealt{1993ApJ...402..593S}). 
This gives:
\begin{equation}
\label{eq3:Rm}
r_{\rm in} \pi\langle S_{z \phi}\rangle/\Omega_* =\dot{m}.
\end{equation}
\bhl [Note that we take the sign of $\dot m$ positive for {\em inward} mass flow.]
 It is not necessary that stationarity holds: (\ref{eq3:Rm}) can also be applied
 when the inner edge of the disc moves. However, since it describes the 
accretion through the magnetosphere-disc boundary $r_{\rm in}$, it has to be 
applied in a frame comoving with $r_{\rm in}$. If $\dot m $ is the mass flow rate
in a fixed frame, it is related to the accretion rate in this comoving frame 
($_{\rm co}$) by
\beq \dot m_{\rm co}=\dot m+2\pi r_{\rm in}\Sigma(r_{\rm in})\,\dot r_{\rm in}, \label{dotm}\eeq
where $\dot r_{\rm in}$ is the rate of change of the inner disc edge.


To connect the accreting case with the dead disc case, for which 
$\dot m=0$, we need one more prescription, this time for the accretion 
rate as a function of the inner edge radius. We introduce a connecting function 
$y_{\rm m}$ for this (DS10):
\beq 
\dot m_{\rm co}(r_{\rm in})=y_{\rm m}(r_{\rm in})\dot m_{\rm a}(r_{\rm in}), \label{dotmm}
\eeq
with the properties $y_{\rm m}\rightarrow 1$ ($r_{\rm c}- r_{\rm in}\gg\Delta r_2$),  
$y_{\rm m}\rightarrow 0$ ($r_{\rm in}- r_{\rm c}\gg\Delta r_2$), with

\begin{equation}
y_m = \frac{1}{2}\left[1 - \tanh\left(\frac{r_{\rm in}-r_{\rm c}}{\Delta r_2}\right)\right],
\end{equation}
where $\Delta r_2$ describes the width of the transition (different in general from 
$\Delta r$). \ehl


\bhl With the star's assumed field of dipole moment $\mu$, $B_{\rm d}=\mu/r^3$ and
 Keplerian orbits in the disc, (\ref{dotm}) becomes\ehl, with the viscous thin-disc 
expression  for $\dot m$:
\begin{equation}
\label{eq3:mdot,final2}
 6\pi r^{1/2}_{\rm in}\ddr(\nu\Sigma r^{1/2 })\big|_{r_{\rm in}}  =   y_m\frac{\eta\mu^2}{4\Omega_*r_{\rm in}^5} - 2\pi r_{\rm in}\Sigma(r_{\rm in})\dot{r}_{\rm in},
\end{equation}
where $\eta$ is a numerical factor of order unity and $\mu$ the dipole 
moment of the star (see DS10 for details).

Along with our description for the viscosity, (\ref{eq3:sigma_rin}) and
(\ref{eq3:mdot,final2}) define a boundary condition at \rin\ and an
equation for \rin(t), for a disc over a continuous range of accretion
rates, from strongly accreting systems ($r_{\rm in} \ll r_{\rm c}$) to
dead-disc systems ($\dot{m} \simeq 0$).

\subsubsection{Evolution of corotation radius}
\label{sec3:dtrc}
In order to study the response of a disc to changes in spin of the
star, we must incorporate the 
angular momentum exchange between the star and disc:
\begin{equation}
I_*\frac{\rd\Omega_*}{\rd t} = \frac{\rd J}{\rd t},
\end{equation}
which introduces the moment of inertia of the star, $ I_* = k^2M_*R^2_*$
as an additional parameter of the problem.

The disc-star angular momentum exchange \bhl $\rd J/\rd t$ has two components: 
matter accreting onto the star adds angular momentum at a rate $\dot{m}_{\rm co} r^2_{\rm in}\Omega_{\rm K}(r_{\rm in})$, while the disc-field coupling outside co-rotation extracts angular momentum spinning the star down. The rate of angular momentum exchange between the
disc to the star will thus be (with \ref{TB}, \ref{BT}, \ref{ys}):
\begin{eqnarray}
\label{eq3:jdot}
  \nonumber \frac{ \rd J}{\rd t} &=& \dot{m}_{\rm co} r^2_{\rm in}\Omega_{\rm K}(r_{\rm in})
- T_B\\
  &=& \frac{\eta \mu^2}{r^3_{\rm in}}\left[\frac{1}{4}\left(\frac{r_{\rm c}}{r_{\rm
        in}}\right)^{3/2}y_m - \frac{\Delta r}{r_{\rm  in}} y_\Sigma\right].
\end{eqnarray}
The corotation radius (a function of $\Omega_*$) \bhl evolves as\ehl:
\begin{equation}
\label{eq3:drcdt}
\frac{\rd r_{\rm c}}{\rd t} = -\frac{2}{3}\frac{\rd J}{\rd t} I^{-1}_*\left(\frac{GM_*}{r^5_{\rm
    c}}\right)^{-1/2}.
\end{equation}

Eq. (\ref{eq3:jdot}) shows that \bhl there is a value of $\dot{m}$ for which there is 
no net angular momentum exchange with the star. This is the `spin equilibrium'
state discussed in previous work. This\ehl\ zero-point will
depend on our adopted connecting functions, as well as the size of the
transition widths, $\Delta r$ and $\Delta r_2$.  \bhl If\ehl\  
$\dot{m} = 0$, there is no spin-equilibrium solution: the \bhl star will\ehl\ 
spin down by the magnetic torque.

\subsubsection{Steady-state solutions}
\label{sec3:steadystate}

\bhl
In the presence of a magnetic torque at the disc inner edge, the steady solutions 
($\partial/\partial t=0$) of the thin viscous disc diffusion equation with the above 
boundary conditions have the form 
(cf.\ DS10):

\beq
\label{eq3:sig}
3\pi\nu\Sigma ={T_B\over\Omega(r_{\rm in})r_{\rm in}^2}\left({r_{\rm in}\over r}
\right)^{1/2}+\dot{m}\left[1-\left(\frac{r_{\rm in}}{r}\right)^{1/2}\right],
\eeq
where $\dot m$ is the accretion rate onto the star, given by (\ref{dotmm}).
If the inner edge is inside corotation ($T_B=0$), 
$\Sigma$ has the standard form for steady accretion on an object rotating below 
the Keplerian rate (second term on the RHS). 

For \rin\ well outside corotation (\rin-\rc $\gg \Delta r$), 
$\dot m\downarrow 0$ and we have a dead disc. The surface density  is then 
determined by the first term on the RHS. The steady 
outward flux of angular momentum in this case has to be taken up by 
a sink at some larger distance, otherwise the disc can not be stationary 
as assumed. This sink can be the orbital angular momentum 
of a companion star, or the disc can be approximated as infinite. The 
latter is a good approximation for changes in the inner regions 
of the disc, if time scales short compared with the viscous evolution of the 
outer disc are considered\ehl.



\subsection{Numerical method}
\label{sec3:numsim}
We use the one-dimensional numerical code described in DS10 to evolve
the standard diffusive thin-disc equation with our viscosity
prescription (\ref{eq3:visc}) and our description of the disc-field
interaction (which gives the inner boundary conditions the boundary
conditions \ref{eq3:sigma_rin} and \ref{eq3:mdot,final2}). At the 
outer boundary a mass flux and a flux of angular momentum are specified 
in various combinations 
(described in sec. \ref{sec3:rout}).

\bhl The calculations are done in dimensionless coordinates and variables. \ehl
In DS10 we scaled all physical lengthscales to \rc, and physical time 
scales to $T_{\rm visc}(r_{\rm c})$. Since in this paper we want to follow the 
evolution of \rc, we instead use the \bhl stellar radius\ehl\ $r_*$ and $T_{\rm
  visc}(r_*)\equiv T_*$ to scale our physical length and timescales.
The grid is logarithmically spaced (to ensure sufficient resolution in the inner
disc to capture the disc instability). \bhl It is an adaptive mesh, such that the inner 
boundary\ehl\ moves with \rin. 

\bhl Since the grid used is time dependent, the outer boundary 
condition is also applied at a time-varying location.  As discussed in 
DS10, the artefacts this causes are small, compared to specification at a fixed
location (at least for the large discs studied in most cases).\ehl

\bhl The size of the discs studied range from $10$ to $10^6$ times
the inner edge radius, and the number of grid points needed for sufficient
resolution varies accordingly, from 90 for the smallest to 560 for
the largest discs.

Time-stepping is done with an implicit method, so that the short
timescales encountered during episodes of cyclic accretion can be
followed, as can the much slower viscous evolution of the disc as a
whole and the spin down the star. It is adapted to the stiff nature of
the equation to be solved (see DS10 for details).\ehl


\subsubsection{Outer Boundary Condition}
\label{sec3:rout}
The lifetime and evolution of a star surrounded by a dead disc is an
inherently time-dependent problem, so the initial conditions in the
disc can be critical for its evolution. Since the spin-down timescale
for the star can be much larger than viscous timescales throughout the
disc, the conditions in the outer disc will also strongly influence
\bhl the evolution of the system.\ehl

We thus consider \bhl the effect of varying the\ehl\  outer
boundary conditions for the disc.
\bhl The first condition we study is the simplest: a fixed mass flux 
$\dot{m} =\dot{m}_0$ ($>0$, corresponding to accretion). 
As discussed in the introduction, a key aspect of disc-magnetosphere 
interaction is that accretion is possible even as the star is spun down.
At fixed $\dot m>0$, the angular momentum flux can be either inward or 
outward. 

If the mass flux specified vanishes at \rout, the boundary condition is \ehl
\begin{equation}
\label{eq3:outbound}
\ddr (r^{1/2}\nu\Sigma)\big|_{r_{\rm out}} = 0.
\end{equation}

On long evolution timescales, the finite extent of the disc itself
could be relevant in a star without a companion where the disc can
spread outwards. To model this, as our final boundary condition we
take $\Sigma(r_{\rm out}) = 0$, so that the angular momentum added at
\rin\ is carried away by the outer parts of the disc, causing the disc
to spread outwards. In section \ref{sec3:const_am} we discuss the
consequences of these assumptions in limiting the lifetime of a
trapped disc. 

\subsubsection{The Evolution of \rc}
The final modification to our code used in DS10 is to allow \rc\ to
evolve as the spin rate of the star changes (\ref{eq3:drcdt}). The
characteristic evolution timescale for $\dot{r}_{\rm c}$, the
spin-down timescale for the star, is much longer than the nominal
viscous timescale in the disc (see the next section). 
The code updates  \rc\  by an explicit time step, 
rather than implicitly, as we do the other variables. 
Rather than discretizing (\ref{eq3:drcdt}) and add it to
our system of linearized equations that are solved numerically at each
timestep, we instead approximate the evolution in \rc\ to first order
in time, that is:
\begin{equation}
r_{\rm c}(t_0 + \Delta t) = r_{\rm c}(t_0) + \frac{\rd r_{\rm c}}{\rd t}\bigg|_{t_0}\Delta t 
\end{equation}
This scheme  is
simpler than adding additional equations to the code, and is
sufficient to describe the co-evolution of the disc and stellar
spin-rate. However, as we discuss in sec. \ref{sec3:ansol},  
\bhl it is not accurate enough\ehl\ when \rin\ is close to
\rc\ \bhl\  and the spindown timescale comparable\ehl\  to the viscous
timescale in the inner regions of the disc.

\section{Characteristic Timescales of Disc-Star Evolution}
\label{sec3:timescale}
 \bhl Three kinds of timescale play a role in the evolution of a star coupled
 magnetically to an accretion disc. These are the spin period of the star, 
 the timescale for changes in spin period of the star, and viscous evolution 
timescales of the disc. The  viscous evolution does not 
have a single characteristic timescale; it can vary over many orders of 
magnitude depending on which regions in the disc participate in the evolution. \ehl

\bhm The spin period of the star ($P_*$) determines the location of
the corotation radius which in turn sets the accretion rate at which
the magnetic torque changes sign. $P_*$ also determines the timescale
for magnetospheric variability (from processes like reconnection of
field lines), which can lead to variability in $\dot{m}$, $\Delta r$,
$\Delta r_2$ and the $B_\phi$ component of the magnetic field (which
sets the magnitude of the torque). $P_*$ is much shorter than the
other characteristic timescales studied in this paper, and the complex
variability processes are best studied with detailed MHD simulations,
so in this work we assume time-averaged values for $\dot{m}$, $\Delta
r$, $\Delta r_2$ and $B_\phi$ and neglect timescales much shorter than
the characteristic viscous timescale at $R_*$. \ehm

\bhl A convenient unit of time for measuring changes in a viscous disc at a 
distance $r$ from the center is $t_{\rm a}=r^2/\nu(r)$, sometimes called 
the accretion- or viscous timescale at distance $r$. If $\alpha$ is the 
viscosity parameter and $H$ the disc thickness, it is longer than the 
orbital timescale $\Omega_{\rm K}^{-1}$ by a factor $\alpha^{-1}(r/H)^2$, 
a large number for most observed discs. Natural choices for $r$ in this 
expression would be the disc's inner edge \rin\ ($t_{\rm in}$) or the
corotation radius \rc ($t_{\rm c}$). 
Both of these are functions of time. The actual timescales of variation in 
our discs can be much shorter than $t_{\rm c}$ and $t_{\rm in}$, however, since the extent 
of the disc that participates in the variation can be much smaller than \rin. 
In the cyclic accretion mode described in DS10, for instance, cycle 
times as short as $0.01 t_{\rm c}$ are found. The timescale for viscous
adjustment in the outer disc regions, on the other hand, can be very large 
compared to $t_{\rm c}$.\ehl

The \bhl longest\ehl\  timescale is the rate at which the star's spin will change,
which is determined both by the rate of angular momentum exchange with
the disc and the star's moment of inertia. The spin-down torque of a
dead disc (with $\dot{m} = 0$ and $r_{\rm in} = r_{\rm c}$) is, \bhl from eq.\ 
(\ref{eq3:jdot}):
\begin{equation}
I_*\frac{{\rm d} \Omega_*}{{\rm d}t} = - \frac{\eta \mu^2\delta}{r^3_{\rm in}},
\end{equation}
where $ \delta={\Delta r}/{r_{\rm in}}$. The characteristic spindown time is:
\begin{equation}
\label{eq3:tsd}
T_{\rm SD} \equiv P_*/\dot P \sim  \frac{I_*\Omega_*r_{\rm in}^3}{\eta\mu^2\delta }.
\end{equation}
If the inner edge stays near corotation, $\Omega_{\rm K}=\Omega_*$, 
replacing \rin\ by \rc, this yields: 

\begin{equation}
\label{eq3:tsd1}
T_{\rm SD} = \frac{GM_*I_*}{2\pi\eta\mu^2\delta}P_*,\quad (\Omega_{\rm K}(r_{\rm in})=\Omega_*).
\end{equation}
The spindown timescale varies considerably between different
sources. Adopting $\eta = 1$ and $\Delta r/r_{\rm in} = 0.3$, this
spindown timescale is short enough to account for spin-regulation in
slowly rotating magnetic stars. In Table \ref{table:tsd} we summarize
the predicted spin-down timescales for a slowly rotating X-ray pulsar,
a millisecond pulsar, a \bhl slowly rotating\ehl\ Ap star, and a \bhl 
typical\ehl\ T Tauri star. \bhl For all these examples\ehl\ but the
millisecond pulsar the spin-down timescale is much shorter than the
lifetime of the star. \bhl Provided the conditions are such that the inner edge 
of the disc can stay near corotation (i.e. what we have called the `trapped 
disc' state), it\ehl\ 
will be able to spin down a star to
very slow rotation periods. In the next sections we will explore how
this could work in detail by evolving a viscous disc in time numerically.

\begin{table*}
\label{table:tsd}
 \centering
    \begin{minipage}{140mm}
      \caption{Spindown and viscous timescales for different type of
        magnetic stars }
    \begin{tabular}{ l | c | c | c | c | l | c | c }
      \hline
        Source & Mass & Radius & $B_*$ & $P_*$ & $I_*$ & $T_{\rm SD}$ & $T_{\rm visc}(r_{\rm c})/T_{\rm SD}$ \\ 
        & ($M_\odot$) &  ($R_\odot$) &(G) & &  ($M_\odot R^2_\odot$) & (years) & \\
        \hline
        slow Pulsar & $1.4$ & $1.4\times 10^{-5}$ & $10^{12}$ & 5.0s &
        $2.9\times 10^{-11}$ & 4400 & $3\times10^{-7}$\\
        ms Pulsar & $1.4$ & $1.4\times 10^{-5}$ & $10^8$ & 0.1s &
        $2.9\times 10^{-11}$ & $2.6\times10^{11}$ & $2\times 10^{-17}$ \\
        Magnetic Ae star\footnote{ $B_*$ and $I_*$ from \cite{2000A&A...353..227S}}
        & $3.0$ & $5.5$ & $10^4$ & 10~yrs & $4.0$ &
        $3\times10^5$ & 0.06\\
        T Tauri Star\footnote{\cite{2009A&A...507..881S}} & 
        $0.6$ & $3.0$ & $1500$ & 7~days & $0.54$ &  $2.3\times10^4$  & 0.001\\  
      \hline
      \end{tabular}
    \end{minipage}
\end{table*}

The last column of Table \ref{table:tsd} lists the ratio of the viscous
timescale $r_{\rm c}^2/\nu$ and (\ref{eq3:tsd}). 
Note that for our description of viscosity, the viscous and spin-down
timescales both scale as $r^{-3/2}$. The quantity $T_{\rm visc}/T_{\rm
  SD}$ thus defines the ratio of the time that gas at that radius
takes to travel inwards onto the star and the time it would take to
spin-down so that $r_{\rm c} = r_{\rm in}$, independent of radius.  In
all cases, the spin-down timescale is much longer than the viscous
timescales in the disc, so that at least part of the disc is able to
adjust to the new spin rate of the star. However, the exact ratio
between the two timescales varies over ten orders of magnitude, from
0.06 for a strongly magnetic Herbig Ae star to $10^{-17}$ for an
accreting millisecond pulsar. This ratio implies that the extent of
spin-down will be influenced by the viscous evolution of the disc
itself in response to the disc-field interaction, and that this
evolution will vary substantially between different systems, breaking
the scale invariance usually assumed in disc-magnetospheric
interactions. In section \ref{sec3:ansol} we demonstrate how the ratio
of these two timescales is critical in determining the ratio of
$r_{\rm in}$ to $r_{\rm c}$ in a trapped disc.

\subsection{Representative Model}
\label{sec3:standard}
In sections \ref{sec3:form_disc} and \ref{sec3:const_am} we study how a
trapped disc can form and evolve, as well as how it can become
untrapped. In order to simplify comparison between different
simulations, we adopt a set of parameters for a representative model,
which we then vary between solutions as necessary. 

For the dimensionless parameters we adopt $\Delta r/r_{\rm in} = 0.1$,
$\Delta r_2/r_{\rm in} = 0.04$, and $B_\phi/B_{\rm z}\equiv
\eta = 0.1$. The values of $\Delta r/r_{\rm in}$ and $\Delta
r_2/r_{\rm in}$ are small enough to provide an abrupt transition
between an accreting and non-accreting disc, but do not show the
cyclic instability discussed in DS10. Neglecting the star's spin
change, the problem has a scale invariance (DS10), whereby the
parameters $\mu$, $M_*$, $\Omega_*$ and $\dot{m}$ can be re-written as
the ratio $\dot{m}/\dot{m}_{\rm c}$, and $\dot{m}_{\rm c}$ is the
accretion rate in (\ref{eq3:Rm}) that puts the magnetospheric radius at
$r_{\rm c}$, \bhl a natural unit of $\dot m$ for magnetospheric accretion. 
The results in DS10 were presented in this unit.\ehl\ 

\bhl As discussed in the previous section, 
the variation of \rc with time during spindown of the star makes this unit 
impractical. Instead, we present the representative model in units 
suitable\ehl\  for a protostellar system with $T_{\rm
   visc}/T_{\rm sd} = 2.6\times10^{-3}$ (which is as large a ratio
as the present version of the code allows), and explore \bhl the effect of 
varying this ratio.
As unit of length we use $r_*$, the star's radius, and for timescale \ehl\  
$t_*$, the nominal viscous timescale of the disc at the star's radius.

\section{Trapped discs}
\label{sec3:form_disc}

\subsection{Trapped disc evolving from an accreting disc}
For our description of the disc-field interaction (which ignores
outflows), once the accretion rate falls to zero, the inner edge of
the disc could be located anywhere outside \rc, depending on the
amount of mass in the disc. What then determines the location of the
inner radius of a dead disc? 
To answer this question, we simulate an initially \bhl steadily\ehl\ accreting
disc in which the accretion rate at $r_{\rm out}$ suddenly decreases
to zero. As the reservoir of gas in the disc runs out, the accretion
rate onto the star declines, and the inner radius of the disc moves
outwards.

In the simulation we use our representative disc parameters described
above, and set the initial inner radius of the disc to be just inside
\rc, $r_{\rm in}(t=0) = 0.88 r_{\rm c,0}$, and the outer radius $r_{\rm out} = 100\, r_{\rm
  in}$. We can calculate the corresponding accretion rate from
(\ref{eq3:Rm}), and use the static solution for $\Sigma$ given by
(\ref{eq3:sig}) as our initial surface density profile.  At $t=0$, we
set $\partial_r(\nu\Sigma r^{1/2})\big|r_{\rm out} = 0$, so that no
mass is added to the disc or allowed to escape. This sets a constant
angular momentum flux at \rout.

\begin{figure}
\rotatebox{270}{\resizebox{!}{84mm}{\includegraphics[width=84mm]{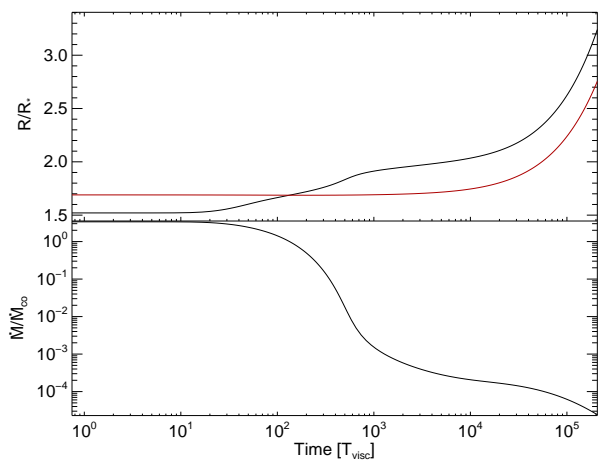}}}
\caption{Transition from an accreting to a \bhl `trapped disc' state\ehl. The
        initial $\Sigma$ profile of the disc sets $r_{\rm in}=1.5 r_*$, but
        the accretion rate through \rout\ is set to zero at $t=0$. 
        \bhl As a result of mass loss by accretion through \rc, \ehl\ the disc quickly 
        evolves away from a steady accretion state \bhl and after about 1500 $t_*$ 
        settles into \bhl a slowly  evolving\ehl\ state in which \rin\ tracks \rc\ehl. Top:
        The black solid curve shows the evolution of the inner radius in
        time, while the red dashed curve shows the evolution of \rc. Bottom:
        The accretion rate onto the star, which decreases sharply as \rin\
        moves outwards \bhl across the corotation radius\ehl.\label{fig3:acc2non}}
\end{figure}

The results are shown in Fig.~\ref{fig3:acc2non}. The bottom panel of
Fig.~\ref{fig3:acc2non} shows the change in accretion rate onto the
star, scaled to $\dot{m}_{\rm c}$.  The top panel of
Fig.~\ref{fig3:acc2non} shows the evolution of the inner radius (solid
black curve) and \rc\ (dashed red curve) in response to the changing
accretion rate. After initial steady accretion over about $30~t_*$
(less than 1/10th \bhl the viscous\ehl\ timescale at \rout), $\dot{m}$
through the inner edge of the disc begins to decrease as the reservoir
of mass in the disc is accreted onto the star, and \rin\ begins to
move outward. From $30 -1500~T_{*}$, $\dot{m}$ decreases
exponentially with a decay timescale of about $240~t_*$ and the disc
moves outwards. However, \rin\ increases by only about 20\% as the
accretion rate decreases by three orders of magnitude. (The structure
of \rin\ around \rc\ is an artifact of the $\tanh$ connecting
functions we adopted to describe \rin\ and $\dot{m}$ across the
transition region). 

After $\sim 1500~t_*$ the star begins to spin down (so that \rc\ moves
outwards), and the inner radius of the disc begins to track \rc, so
that the ratio $r_{\rm in}/r_{\rm c}$ remains nearly constant
thereafter. The behaviour of the accretion rate at this point also
changes. Although it continues to decrease exponentially, the decay
timescale lengthens considerably and the accretion rate ($\sim
10^{-4}$ of the initial $\dot{m}$) is regulated by the spin-down rate
of the star. Instead of continuing to move away from \rc\ into the
`dead disc' regime (in which $\dot{m}=0$), the inner radius instead
remains {\em trapped} at nearly a constant fraction of \rc\ while the
star continues to spin down. We thus call this disc solution a
`trapped disc'', since rather than continue to move outwards, \rin\
becomes trapped at a nearly constant fraction of \rc. 

At the outer edge of the disc $\dot{m} = 0$ and \bhl there is an outward 
angular momentum flux\ehl. The accretion onto the star \bhl comes\ehl\  
from the disc being slowly eroded  (although at a very low 
rate) \bhl as \rin\ moves outward\ehl.  The evolution of ${r}_{\rm in}$ and the 
inner parts of the disc  is determined by the spin down rate of the
star itself, which is itself influenced by how close \rin\ can \bhl stay near\ehl\ 
\rc. In Sec. \ref{sec3:ansol} we demonstrate how $r_{\rm
   in}/r_{\rm c}$ is mostly determined by the parameters $\Delta r$
and $\Delta r_2$, and the ratio $T_{\rm visc}/T_{\rm
    SD}$. However, the main conclusion of this section is clear: if a
trapped disc forms and can efficiently carry away the angular momentum of
the star, over spin-down timescales the disc will accrete at such a
rate so as the inner edge of the disc can move outwards together with
\rc\, and the star could in theory spin down \bhl completely\ehl.

\subsection{Trapped disc evolving from a dead disc}
\label{sec3:dead}

\bhl Consider next a case where the initial condition is a dead disc, 
(\rin~$>$~\rc) given by the steady profile (\eqref{eq3:sig} with $\dot m=0$ 
and $y_\Sigma=1$). As the star spins down, \rc\ moves out until it 
catches up with the inner edge \rin.
From then on, the same evolution is as in the previous case: \rin\ and 
\rc\ move outward together indefinitely. A small amount of 
mass accretes while the star's angular momentum is transferred 
to the disc.

The asymptotic evolution of this dead disc can be compared with the
case when a fixed mass flux is imposed at the outer edge. The 
asymptotic state is then a steady state with {\em spin equilibrium}: 
the spin-up torque due to the accreted mass is balanced by the
magnetic torque at \rc\ transfering angular momentum outward.\ehl

\begin{figure}
  \rotatebox{90}{\resizebox{!}{84mm}{\includegraphics[width=84mm]{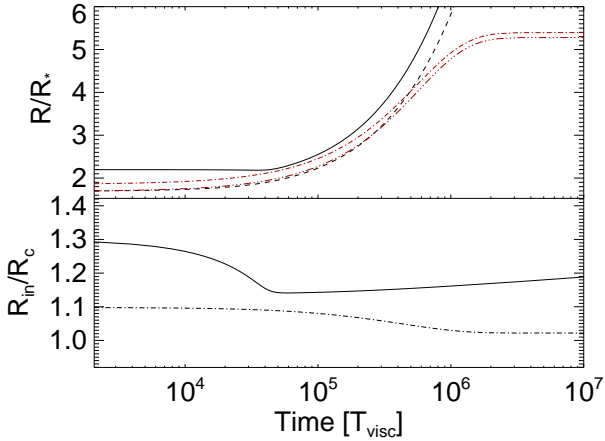}}}
     \caption{Comparison of the evolution of \rin\ and \rc\ between a dead disc 
        (black solid and dashed curves) and accreting disc (red dot-dashed curves). 
        Top: The evolution of \rc\ and \rin\ in units of the stellar radius.
        Black curves: \rin\ (solid) and \rc\
        (dashed) in the dead disc;  red curves:  the accreting disc (\rin,
        dot-dashed curve; \rc, triple-dot dashed curve). 
        Bottom: The ratio $r_{\rm in}/r_{\rm c}$ for the dead disc (solid) and 
        the accreting disc (dot-dashed). 
        The dead disc keeps evolving indefinitely, the accreting case reaches 
        a steady state in spin equilibrium with the star around $t\sim 10^6$.
        \label{fig3:dead}}
 \end{figure}

We illustrate the distinction between these two \bhl cases\ehl\ in
figure \ref{fig3:dead}. This shows the evolution of an initially dead
disc (black, thick curves) and accreting disc (red, thin curves). Both
discs have the same representative parameters
(Sec. \ref{sec3:standard}) and $r_{\rm out} = 100\,r_{\rm in}$, but
with different initial inner radii \bhl(a few times the stellar
radius)\ehl, accretion rates and appropriate initial surface density
profiles given by (\ref{eq3:sig}). For the dead disc, we take $r_{\rm
  in} = 1.3 r_{\rm c,0}$ (where $r_{\rm c,0}$ is the initial
corotation radius) which corresponds to $\dot{m} \simeq 0$ for our
chosen value of $\Delta r_2/r_{\rm in}$. For the accreting disc, we
choose $r_{\rm in} = 1.1 r_{\rm c,0}$, which corresponds to a low but
non-zero accretion rate ($\dot{m} = 8\times 10^{-3} \dot{m}_{\rm c}$).

Since $\dot{m}$ for the accreting
disc is initially low compared to $\dot{m}_{\rm c}$, at early times
\rc\ evolves at the same rate in both simulations (accreting: red,
triple-dashed curve; non-accreting: black, dashed curve), and the star
spins down. Eventually, however, the amount of angular momentum added
by the accreted gas becomes comparable to the amount removed, 
\bhl and spin equilibrium is reached\ehl\ at $\sim 10^6~t_*$. In 
contrast, for the $\dot{m} = 0$ case, the disc at first remains \bhl 
unaffected\ehl, while \rc\ moves outwards. \bhl(This is because the 
magnetic torque depends only on distance, not on the rotation rate 
of the star)\ehl. After \rc\ moves close enough
to \rin\ that accretion can begin (around $4\times 10^4~t_*$), the two
start to move outwards at approximately the same rate.  The \bhl(low)\ehl\
accretion rate onto the star is determined by the \bhl(slow)\ehl\ rate at which \rc\
moves outwards, and the star continues to spin down indefinitely. The
bottom panel of fig. \ref{fig3:dead} shows the ratio of $r_{\rm in}/r_{\rm c}$ 
for the accreting (dashed) and  non-accreting 
disc (solid). After the non-accreting disc passes out of the dead disc
phase (at $\sim 4\times 10^4$), in both systems the ratio changes by
less than 10\%, and \rin\ always remains close to \rc. \bhl The 
non-accreting disc, however, differs in that it\ehl\ never reaches spin 
equilibrium.

The main difference between the evolution of a dead disc and an
accreting disc is the behaviour of the inner \bhl edge radius\ehl. As 
seen in sec. \ref{sec3:dead}, in \bhl the initially\ehl\ dead disc the accretion 
rate onto the star is determined by the disc's behaviour when \bhl it reaches\ehl\ $r_{\rm in} 
\simeq r_{\rm c}$, while in an accreting disc the accretion rate is 
\bhl governed by the value set at the outer boundary\ehl. 


\bhl Both the accretion rate on the star and the outward angular momentum 
flux in our trapped discs depend sensitively on the distance between 
\rin\ and \rc\ compared with the transition widths $\Delta r$ and $\Delta r_2$. 
In the results of figure \ref{fig3:dead}, \rin - \rc\  is of the order $0.5-2\, 
\Delta r$. In the next section we develop an analytic estimate of this 
number and compare it with the numerical results.\ehl


\subsection{Analytic estimates for a trapped disc }
\label{sec3:ansol}

As we showed above, an initially dead disc will eventually start
accreting at a low rate, \bhl in such a way that\ehl\ \rin\ moves outwards 
together with \rc\ at a nearly constant ratio. The accretion rate onto the 
star is determined by how close \rc\ can move to \rin\ before the disc 
moves outwards in response. \bhl The actual distance on which \rin\ settles 
in cases like those show in the previous section
depends\ehl\ on the details of the disc-field interaction (namely the
parameters $\Delta r$ and $\Delta r_2$) and the ratio of timescales,
$T_{\rm visc}/T_{\rm SD}$. 


The spin-down timescale derived above assumes that \rin\ moves
steadily outwards at the same rate as \rc. This timescale is an upper
limit, since as \rc\ approaches \rin\ there is reduced transport of
angular momentum through \rin\ and accretion onto the star begins. In
addition, in order for spin-down to remain efficient, the angular
momentum added by the disc-field interaction can be transported
through the disc and carried away at \rout, otherwise \rin\ will move
quickly away from \rc\ and spin-down will effectively cease.

When \rin\ is far enough from \rc\ that $\dot{m} \simeq 0$, \rin\ stays
fixed as the star is spun down and \rc\ moves outwards
(\ref{eq3:sig}). However, once \rin\ moves closer to \rin\ (within
$\Delta r$ or $\Delta r_2$), this static state is no longer possible:
either matter at \rin\ starts accreting onto the star ($y_m \neq 0$),
or the surface density at \rin\ declines ($y_\Sigma \neq 1$), which
causes \rin\ to move closer to \rc\ until accretion through \rin\ can
begin. Since the viscous timescale in the inner part of the disc is
much shorter than the spindown timescale, after accretion through
\rin\ begins, a pseudo-steady-state develops, and \rin\ moves slowly
outwards with \rc. If the disc can maintain a steady-state for the
given $\dot{m}$, then \rin\ will track \rc, and the disc will remain a
nearly dead disc as the star spins down to a small fraction of its
initial spin period.

We can study this quantitatively by considering the equations for
$\dot{r}_{\rm in}$ and $\dot{r}_{\rm c}$. The evolution of the inner
edge of the disc (\ref{eq3:mdot,final2} from Sec. \ref{sec3:dtrin}),
defining $u \equiv \Sigma r$ for convenience is:
\begin{equation}
\label{eq3:dtdrin}
2\pi u(r_{\rm in})\dot{r}_{\rm in} =
y_m\frac{\eta\mu^2}{4\Omega_*r_{\rm in}^5} - 6\nu_0\pi r^{1/2}_{\rm
  in}\frac{\partial u}{\partial r}\big|_{r_{\rm in}}.
\end{equation}

As long as the two terms on the right of (\ref{eq3:dtdrin}) balance,
$\dot{r}_{\rm in}=0$ even after accretion through \rin\ begins. This
will continue until the surface density profile near \rin\ no longer
satisfies (\ref{eq3:sig}). Since the change in \rc\ is the only source
of variability in the problem, \rin\ will approximately track \rc.

\bhl Eq.\ (\ref{eq3:dtdrin}) cannot be solved as is, since $\partial 
u/\partial r$ depends on solving the full time dependent diffusion 
problem. As an estimate we assume that  \ehl\ 
$u$ changes as a result of the changing boundary condition 
(which will increase the surface density gradient), divided by the 
rate at which the rest of the disc can respond to that change (which 
will smooth it out). This can be approximated by:
\begin{equation}
\frac{\partial u}{\partial r}\big|_{r_{\rm in}} \sim -\frac{\partial
  u(r_{\rm in})}{\partial t} v^{-1}_{\rm visc},
\end{equation}
where $v_{\rm visc}$ is the viscous speed at \rin, of order:
$v_{\rm visc} \sim \nu/r$. 

The time derivative for $u(r_{\rm in})$ follows from the boundary
condition for $u$:
\begin{equation}
\frac{\partial u}{\partial t}\big|_{r_{\rm in}} = \frac{\partial u}{\partial r_{\rm in}}
\dot{r}_{\rm in} + \frac{\partial u}{\partial r_{\rm c}}\dot{r}_{\rm c}.
\end{equation}

The equation for $\dot{r}_{\rm in}$ then becomes:
\begin{equation}
\label{eq3:ev_rin}
\dot{r}_{\rm in} = \left(\dfrac{\nu_0 y_m f^{3/2}}{8 r_{\rm in}\partial_{r_{\rm
        in}}y_\Sigma}\left(\dfrac{\Delta r}{r}\right)^{-1}r^{-1/2}_{\rm in} - \dot{r}_{\rm
    c}\right)\left({\dfrac{10}{3}\dfrac{y_{\Sigma}}{\partial_{r_{\rm in}}y_\Sigma
    r_{\rm in}} - 1}\right)^{-1},
\end{equation}
where $f \equiv (r_{\rm c}/r_{\rm in})$, and we have used the
definition of $y_\Sigma$ from Sec. \ref{sec3:sigma_rin} so that
$\partial_{r_{\rm in}}y_\Sigma = -\partial_{r_{\rm c}}y_\Sigma$. 

The evolution of $\dot{r}_{\rm c}$ depends on the rate of angular
momentum exchange with the star. Matter falling onto the star spins it
up, while the interaction with the disc outside \rc\ transfers angular
momentum outward and spins the disc down. The equation for this
evolution is given by (\ref{eq3:drcdt}), which can be re-written:
\begin{equation}
\label{eq3:ev_rc}
\dot{r}_{\rm c} =  \frac{2}{3}\frac{\eta\mu^2}{(GM_*)^{1/2} I_*}f^{7/2} \left[\frac{\Delta r}{r_{\rm in}}y_\Sigma - \frac{y_m}{4}f^{1/2}\right]r^{-1/2}_{\rm in}.
\end{equation}

We can study the evolution of $\dot{r}_{\rm c}$ and $\dot{r}_{\rm
  in}$ in two limiting cases. In the limit where $r_{\rm in} - r_{\rm
  c} \gg \Delta r, \Delta r_2 $:
\begin{eqnarray}
\dot{r}_{\rm c} &\to& \frac{2}{3}\frac{\eta\mu^2 r^{-4}_{\rm
      in}}{(GM_*)^{1/2}I_*}\frac{\Delta r}{r_{\rm in}}r^{7/2}_{\rm c}\\
\nonumber \dot{r}_{\rm in} &\to& 0,
\end{eqnarray}
so that:
\begin{eqnarray}
\label{eq3:ansol_big}
r_{\rm c} &=& \left(-\frac{5}{3}\frac{\eta\mu^2 r^{-4}_{\rm
    in}}{(GM_*)^{1/2}I_*}\frac{\Delta r}{r_{\rm in}} t + r^{-5/2}_{\rm
  c,0}\right)^{-2/5} \\
\nonumber r_{\rm in} &=& r_{\rm in,0}.
\end{eqnarray}

This is the limiting `dead disc' case, where the amount of angular
momentum being injected at \rin\ can be extracted at $r_{\rm out}$ and
the disc remains steady while the star is spun down. It predicts a
slightly smaller spin-down torque than was estimated in
Sec. \ref{sec3:timescale} because the torque scales with \rin.

The inner radius of the disc will remain approximately constant until
either $\dfrac{y_m}{y_\Sigma}$ or $\dfrac{\partial_r
  y_\Sigma}{y_\Sigma}\left(\dfrac{\Delta r}{r_{\rm in}}\right)^{-1}$ become
non-negligible (\ref{eq3:ev_rin}).  Based on our assumption that the
disc will remain in a quasi-steady-state while \rin\ moves, \rin\ will
evolve only in response to changes in \rc, which means that $f$ is a
constant. This simplifies (\ref{eq3:ev_rin}) and (\ref{eq3:ev_rc})
considerably:
\begin{eqnarray}
\label{eq3:pde}
\dot{r}_{\rm c} &=& (A_0 f^{7/2} - A_1 f^{4}) r^{-1/2}_{\rm in}\\
\nonumber \dot{r}_{\rm in} &=& \frac{B_0f^{3/2} r^{-1/2}_{\rm
    in}-\dot{r}_{\rm c}}{B_1 - 1},
\end{eqnarray}
where,
\begin{eqnarray}
A_0&=&\frac{2}{3}\frac{\eta\mu^2}{(GM_*)^{1/2}I_*}\frac{\Delta r}{r_{\rm in}}y_\Sigma\\
\nonumber A_1 &=&\frac{1}{6}\frac{\eta \mu^2}{(GM_*)^{1/2}I_*}y_m\\
\nonumber B_0 &=&\frac{\nu_0 y_m}{8 r_{\rm in} \partial_{r_{\rm
      in}}(y_\Sigma)}\left(\dfrac{\Delta r}{r_{\rm in}}\right)^{-1}\\
\nonumber B_1 &=&\frac{10 y_\Sigma}{3 r_{\rm in}\partial_{r_{\rm in}}(y_\Sigma)}. 
\end{eqnarray}

The solution is then:
\begin{eqnarray}
\label{eq3:ansol}
r_{\rm c} &=& \left(\frac{3}{2}\left(A_0f^3-A_1f^{9/2}\right)t + r^{3/2}_{\rm c,
  0}\right)^{2/3}\\ 
\nonumber r_{\rm  in}&=&\left(\frac{3}{2} \left(\frac {B_0f^{3/2} - A_0
f^{7/2} + A_1 f^4}{B_1-1}\right)t + r^{3/2}_{\rm in}\right)^{2/3}.
\end{eqnarray}
We can use (\ref{eq3:pde}) to calculate $f$, that is, how close \rc\
can move towards \rin\ before the disc will start moving outwards in
response. Setting $\dot{r}_{\rm c} = f\dot{r}_{\rm in}$, we can
re-express the constants in (\ref{eq3:ansol}) as:
\begin{eqnarray}
\label{eq3:fsol}
\left(2\dfrac{\Delta r}{r_{\rm in}} f y_{\Sigma} - \frac{y_m
    f^{3/2}}{2}\right)\left(\frac{10y_\Sigma}{3r_{\rm
      in} \partial_{r_{\rm in}}y_\Sigma} + f -
  1\right)\\
\nonumber\left(\frac{r_{\rm in}\partial_{r_{\rm in}}y_\Sigma}{y_{\rm m}}\right) =
\frac{3}{8}\frac{T_{\rm SD}}{T_{\rm visc}},
\end{eqnarray}
and solve $f$ numerically to give the approximate evolution for \rc\
and \rin\ in time. 

\begin{figure}
\rotatebox{270}{\resizebox{!}{84mm}{\includegraphics[width=84mm]{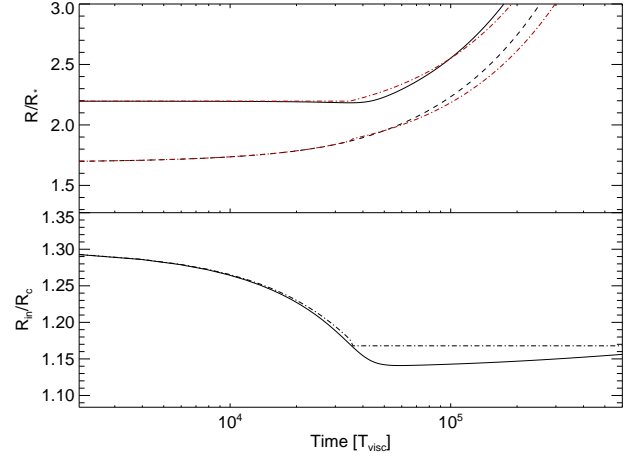}}}
     \caption{Comparison between numerical solution and analytic estimate
        for an idealized dead disc around a star spinning close to
        break-up \rc = 1.7, with $\Delta r/r_{\rm in} = 0.1 $ and $\Delta
        r_2/r_{\rm in} = 0.05$. Top: Numerical solution of the evolution
        of \rc\ (black, dashed curve) and \rin\ (black, solid curve) in
        time. Overplotted is our analytic estimate for \rin\ (red
        dash-dotted curve) and \rc\ (red dash-triple dotted curve) for the
        same physical conditions. Bottom: The ratio \rc/\rin\ for our
        numerical solution (solid curve) and analytic estimate
        (dash-dotted curve). The disagreement in the two solution arises
        from our simplified treatment of the viscous
        disc. \label{fig3:ansol}}
\end{figure}

In Fig.~\ref{fig3:ansol} we compare our estimates for the evolution of
\rc\ and \rin\ to the solution from our numerical simulation. We
consider a rapidly spinning star ($r_{\rm c} = 1.7$), with $\Delta
r/r_{\rm in} = 0.1$ and $\Delta r_2/r_{\rm in} = 0.05$.  In the top
panel we compare the two solutions for \rin\ (estimate: dot-dashed red
curve vs. numerical solution: black solid curve) and \rc\ (estimate:
triple-dotted-dashed red curve vs. numerical solution: black dashed
curve). In the bottom panel we plot the ratio $r_{\rm in} / r_{\rm c}$
for the numerical (solid curve) and analytic (dashed curve) result. At
early times, $r_{\rm c} < r_{\rm in}$ (\ref{eq3:ansol_big}), and the
solutions match exactly. However, at late times the solutions disagree
somewhat. Most obviously, the value for $f$ calculated by
(\ref{eq3:ansol}) is smaller than the numerical solution, that is that
$\dot{r}_{\rm in} = 0 $ for longer than we predict, and there is some
evolution in $f$ over long timescales. This mismatch comes from our
simplified treatment for $\partial_ru\big|_{r_{\rm in}}$, which over
long timescales will depend on the surface density gradients in the
entire disc.

By comparing the sizes of each term on the left-hand size of
(\ref{eq3:fsol}), some insightful approximations can be made. Since the
solution is nearly a dead disc, the disc's accretion rate will be very
low, so that $y_{\rm m} \ll 1$, $y_{\Sigma} \sim O(1)$, and
$\partial_{r_{\rm in}}y_\Sigma \ll 1$. As well, \rin\ is close to \rc\
so that $f\sim O(1)$. Since in general $T_{\rm visc}/T_{\rm
    SD}\ll 1$, (\ref{eq3:fsol}) can be approximated:
\begin{equation}
\label{eq3:fsol_approx}
\dfrac{160}{9}\dfrac{\Delta r}{r_{\rm in}} f \dfrac{y^2_\Sigma}{y_m}
\simeq \dfrac{T_{\rm SD}}{T_{\rm visc}},
\end{equation}
and used to estimate $f$. The left-hand side of
(\ref{eq3:fsol_approx}) is dominated by $y^{-1}_m$, which quickly grows
as \rin\ moves away from \rc, and must balance the right-hand
side. The ability to sustain a dead disc will thus depend on {\em
  both} the interaction between the disc and the field (through the
parameter $\Delta r_2$), {\em and} the ratio between the spin-down
timescale and the accretion timescale.

For a given $T_{\rm SD}/T_{\rm visc}$, if $\Delta
  r_2/{r_{\rm in}} \ll 1$, then $y_{\rm m}$ is nearly a step function,
and $f$ will stay close to 1 even if $T_{\rm SD}/T_{\rm
    visc}$ is very large. On the other hand, if $\Delta
  r_2/r_{\rm in} \sim 1$, and accretion continues even when \rin\
moves a fair distance from \rc, then it is possible that the solution
to (\ref{eq3:fsol}) will predict a larger value for \rin\ than can
support a dead disc. In this case no dead disc solution exists: even
if the disc initially begins as a dead disc, once \rc\ moves close
enough to \rin\ that accretion begins, the disc will begin to move
outwards until matter at \rin\ can be expelled from the system. The
outflow of material will then proceed at a moderate rate as the
surface density profile of the disc evolves away from the dead disc
solution [(\ref{eq3:sig}) with $\dot{m}=0$] to the standard disc solution,
with outflow rather than accretion onto the star.
 
\section{Trapped and untrapped}
\label{sec3:trapuntrap}

\subsection{Accreting discs evolving to trapped or dead disc states}
\label{sec3:trap_dead}
\bhl The results found so far and in DS10 indicate the strong tendency 
for the inner edge of the disc to track the corotation radius, what we
call here a `trapped disc'. This does not happen in all cases, however. 
We would like to find out under what conditions a disc gets stuck 
in this way, and when instead the inner edge proceeds to move 
well outside corotation into the `dead disc'  state.

Armed with qualitative understanding from the previous section, we 
can address this question with a few numerical experiments. We take the
case of a neutron star with field strength $B_{\rm S} = 10^{12}$,
initial spin period $P_*=5{\rm s}$, and
initial inner edge radius $r_{\rm in} = 0.95r_{\rm c,0}$. The disc is thus initially in an accreting state.
The (initial) outer boundary is located at $r_{\rm out} = 100 r_{\rm
  in}$, and we set $\dot{m} = 0$ there so that no matter can escape
the system. The other parameters are the same as for our
representative model.

\begin{figure*}
  \rotatebox{90}{\resizebox{!}{168mm}{\includegraphics[width=168mm]{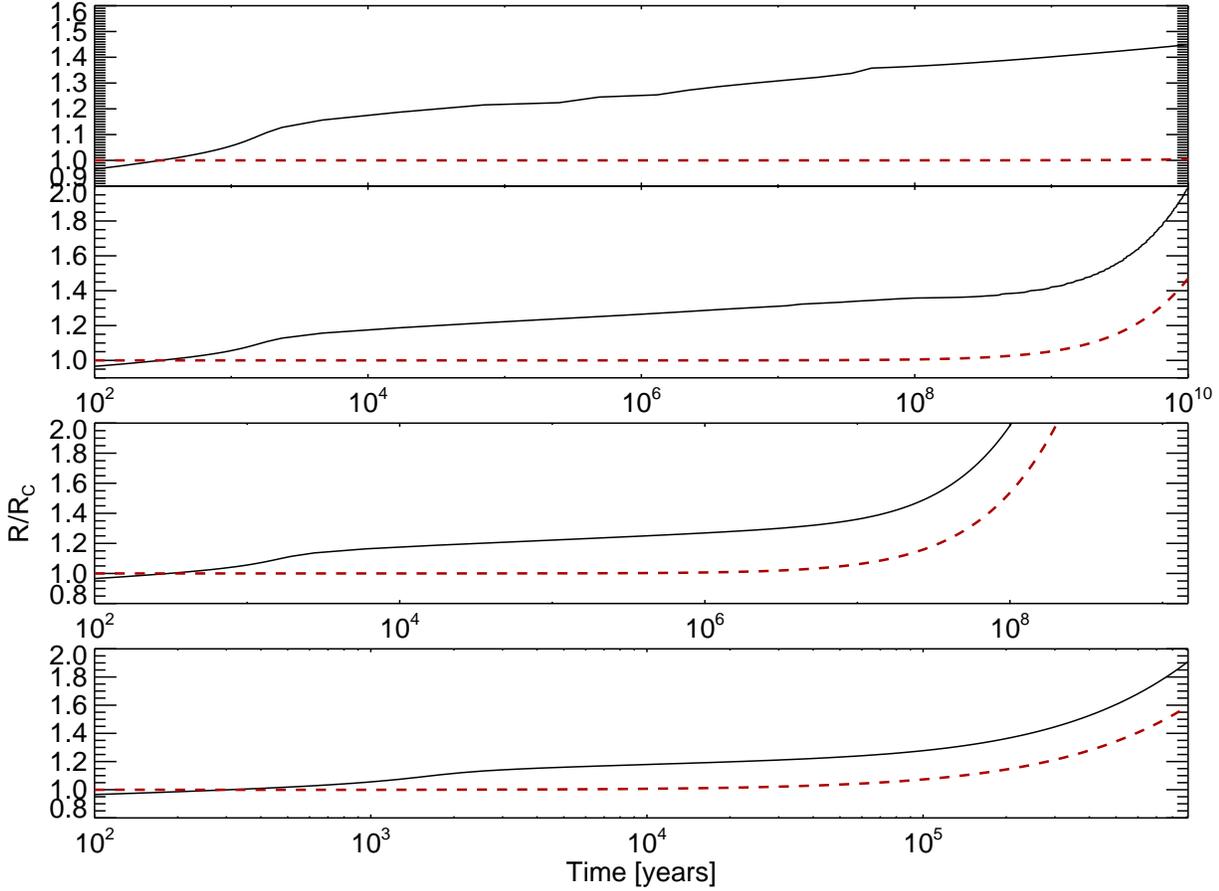}}}
    \caption{The evolution from   an accreting to a non-accreting disc, for increasing (top to
      bottom) ratios of $T_{\rm visc}/T_{\rm SD}$ (with $\Delta r/r_{\rm in}
      = 0.1$, $\Delta r_2/r_{\rm in} = 0.04$). From top to bottom, the
      ratio $T_{\rm visc}/T_{\rm SD}$ is $2.5\times
      [10^{-9},10^{-7},10^{-5},10^{-4}]$. As the ratio between the two
      timescales decreases, the disc is not able to move outwards as
      quickly before the star begins to spin down, so that \rin\ will
      always remain close to \rc. Black solid curve: evolution of
      \rin. Red dashed curve: evolution of \rc.\label{fig3:stable_visc}}
\end{figure*}

We first investigate the effect of varying the viscosity on the way the 
transition from an accreting to a dead disc takes place.  This is shown
in fig. \ref{fig3:stable_visc}. The spindown timescale \eqref{eq3:tsd1}
for the initial \rc\ is $T_{\rm SD} = 10^5$ years. From
top to bottom, we plot this transition for decreasing values of viscosity 
(and hence increasing viscous timescales). To compare these two
quantities we define $T_{\rm visc}$ as the viscous timescale at the
initial \rc. In the different plots, the ratio $T_{\rm visc}/T_{\rm
  SD}$ increases from $2.5\times10^{-9}$ to
$10^{-4}$. 

The most striking change in behaviour occurs between the top panel and
the second panel. At the shortest viscous timescale, the disc does
not get into a trapped state, but evolves directly through corotation,
while at lower viscosty the it always settles into the trapped state,
with the inner edge moving in step with with the corotation radius.

In all the discs, the initial evolution is the same: as $\dot{m}$
decreases, \rin\ moves outwards, crossing \rc\ over about $10^3$
years. Once that happens, however, the subsequent evolution of the
disc and star differes substantially between simulations. In the most
viscous discs (top), \rin\ continues moving steadily outwards over
$10^{10}$ years -- roughly $10^5$ times longer than the nominal
spin-down timescale of the disc. Since \rin\ keeps moving outwards,
the torque on the star decreases too, so that the disc moves far away
from \rc\ before it is able to spin down the star. 

As the viscosity \bhl is reduced\ehl, the ratio between the two
timescales becomes smaller, and \rin\ does not move so far away from
\rc\ before $\dot{r}_{\rm in} \sim \dot{r}_{\rm c}$. Thus for the disc
with the lowest viscosity (bottom), \rin\ and \rc\ begin to move
outwards after about $10^5$ years, and the trapped disc (where
$\dot{m}$ is regulated by $\dot{r}_{\rm c}$) has a much larger
accretion rate onto the star (seven orders of magnitude larger after
$10^4$ years) than for higher viscosities.

\bhl 
By the initial condition chosen, the magnetic torque pushes the inner 
edge out across corotation. This causes mass to pile up outside \rin. 
The higher the viscosity, the faster this pile is reduced again by outward 
spreading, and the faster the inner edge can continue to move outward in 
response. The experimental result is then that trapping behavior is
avoided when the transition takes place fast enough. We return to this
in the discussion.

The pile up is also influenced by the way in which accretion
on the star changes as \rin\ crosses \rc, hence we expect that the parameter
controlling this, $\Delta r_2$, will have a strong effect as well. 
\ehl

\begin{figure*}
  \rotatebox{90}{\resizebox{!}{168mm}{\includegraphics
      [width=168mm]{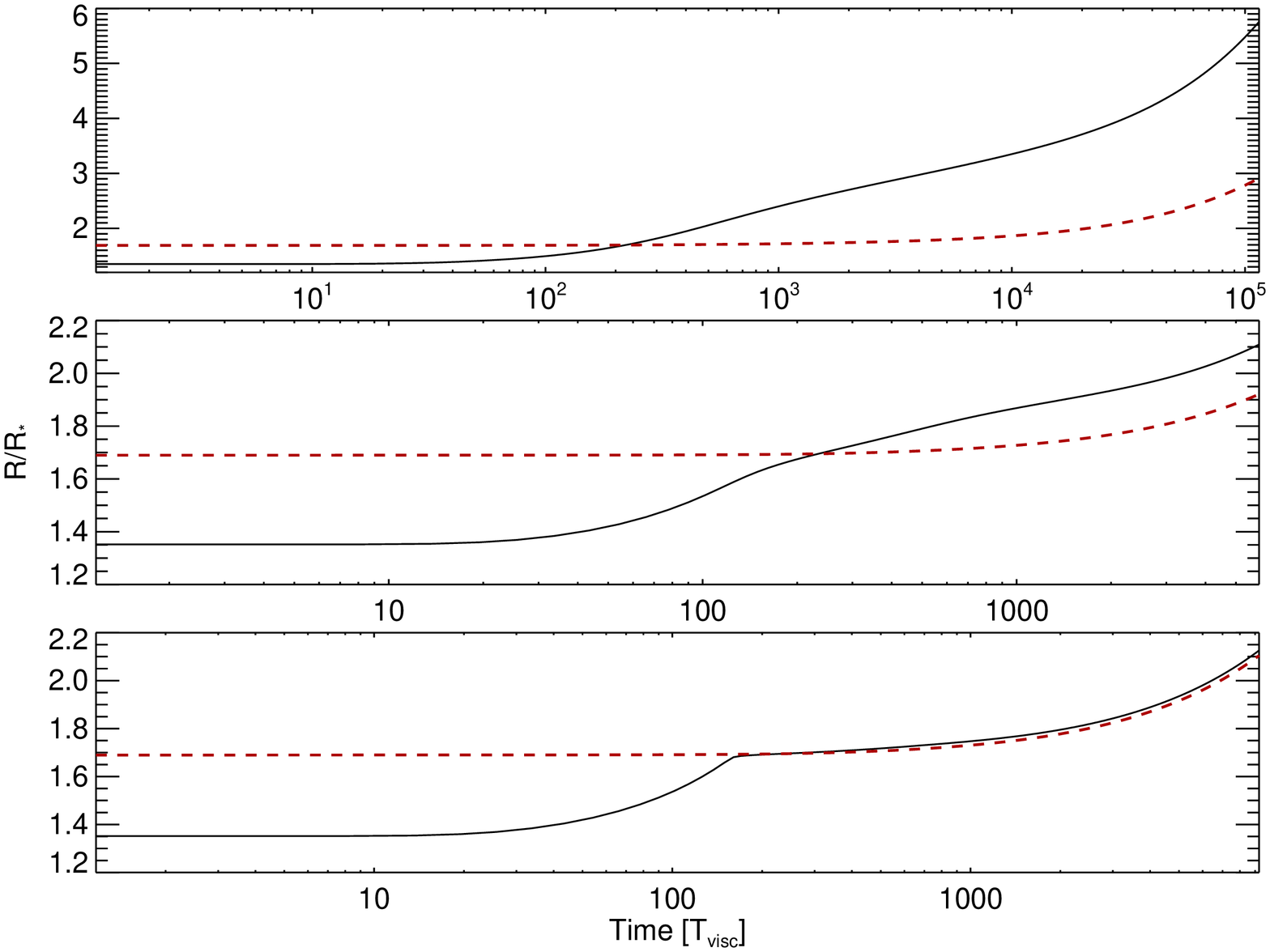}}}\caption{The evolution from
   an accreting to a non-accreting disc, for stable discs ($\Delta
    r/r_{\rm in} = 1$) with different $\Delta r_2$.  \bhl From top to bottom\ehl,
     $\Delta r_2/r_{\rm in} =
    [0.4,0.04,0.004]$. Curves show \rin\ and \rc\ as in fig. \ref{fig3:stable_visc}. \label{fig3:stable_deltar2}}
\end{figure*}

In fig.~\ref{fig3:stable_deltar2} we show 
three discs in which the initial ratio $T_{\rm SD}/T_{\rm visc}$ is
kept fixed, but the value of $\Delta r_2$ changes, from top to bottom,
$\Delta r_2/r_{\rm in} = [0.4,0.04,0.004]$. The larger $\Delta r_2$,
the further away \rin\ must move from \rc\ in order for the accretion
rate to decrease sufficiently to form the trapped disc. In the top
panel (when $\Delta r_2$ is largest), the disc must move out a
considerable distance before becoming trapped, sufficiently far to
significantly decrease the efficiency of the spin-down torque (and
hence increase the spin-down timescale of the star). As well as
decreasing the spin-down efficiency, if $\Delta r_2$ is large enough,
the inner radius could move far enough away from \rc\ that material
could begin to be launched from the disc in an outflow, and the disc
would become untrapped.

As was shown in Section \ref{sec3:timescale}, the ratio $T_{\rm
    visc}/T_{\rm SD}$ itself can vary over many orders of magnitude
in different systems, from $10^{-17}$ in neutron stars with weak
magnetic fields to $10^{-2}$ in discs around massive young stars. The
size of this ratio will also determine whether a trapped disc can
form. 

This analysis would suggest that, assuming $\Delta
  r_2/r_{\rm in}$ does not vary much from system to system, trapped
discs are much more likely to form in protostellar discs than in
strongly ionized discs around neutron stars. Furthermore, the closer
\rin\ is to \rc, the higher the accretion rate in the trapped disc
disc. Conceivably, especially if the viscosity in the disc were very
low, this accretion rate could be larger than the average accretion
rate in the disc itself, so that the disc could spin down the star for
a long time without ever reaching spin equilibrium, even with a finite
accretion rate onto the star.


\bhl As the results reported above show, the evolution can end either
in a dead disc state in which the star has lost only a fraction of its
angular momentum, or a trapped state in which corotation is maintained
and the star can spin down much further. Which outcome results depends
details of the interaction between the star and the disc, parametrized
in our model by the transition widths $\Delta r$ and $\Delta r_2$. It
also depends on\ehl\ the rate at which the disc can respond \bhl
viscously compared to the spin change rate of the star. A fast
response of the disc makes the transition through corotation faster
than \rc\ changes, and the disc is more likely to enter the dead
state. This makes it far more likely to occur in young stellar systems
than in \bhl neutron star binaries\ehl.
\bhl In the results presented above, the initial conditions were 
taken from steady solutions of the viscous thin disc equation.
These  included dead discs in which a steady state was made possible
by a sink of angular momentum at the outer boundary of the numerical 
grid, which takes up the angular momentum added by the magnetic 
 torques at the inner edge\ehl.

\subsection{Dead discs evolving into trapped discs}
\label{sec3:const_am}
In this example we investigate the opposite case of section
\ref{sec3:trap_dead}: discs with \rin\ initially outside corotation, and
conditions chosen such that the disc begins by spreading inward. The
mass flux at the outer boundary is set to zero. Varying the initial
outer radius, $r_{\rm out,0}$ varies the amount of mass in
it, and the timescale of its long-term evolution can change.

\begin{figure}
\rotatebox{90}{\resizebox{!}{84mm}{\includegraphics[width=84mm]{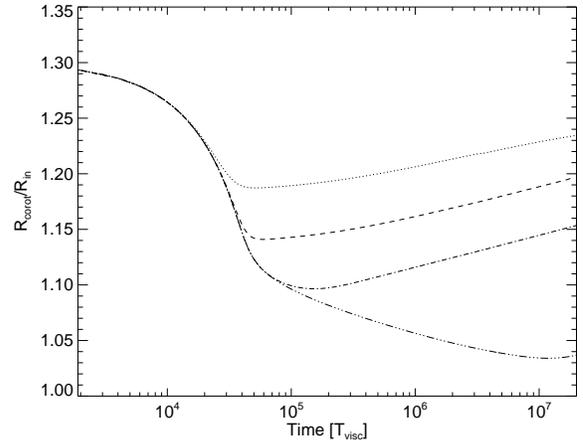}}}
\caption{\bhl Discs starting outside corotation and spreading inward\ehl. 
 Evolution of $r_{\rm in}/r_{\rm c}$ for different initial\rout. From top to
  bottom: $r_{\rm out,0} = 10 r_{\rm in,0}$ (dotted curve), $10^2$ (dashed curve),
  $10^3$ (dash-dotted curve), and $10^4$ (triple dash-dotted curve)
  \rin\ at $t=0$. Larger discs have a larger reservoir of matter, so
  that they can sustain a larger $\dot{m}$ (so smaller \rin) as
  \rc\ increases \bhl due to spindown of the star\ehl. \label{fig3:const_ang1}}
\end{figure}

\begin{figure}
\rotatebox{90}{\resizebox{!}{84mm}{\includegraphics[width=84mm]{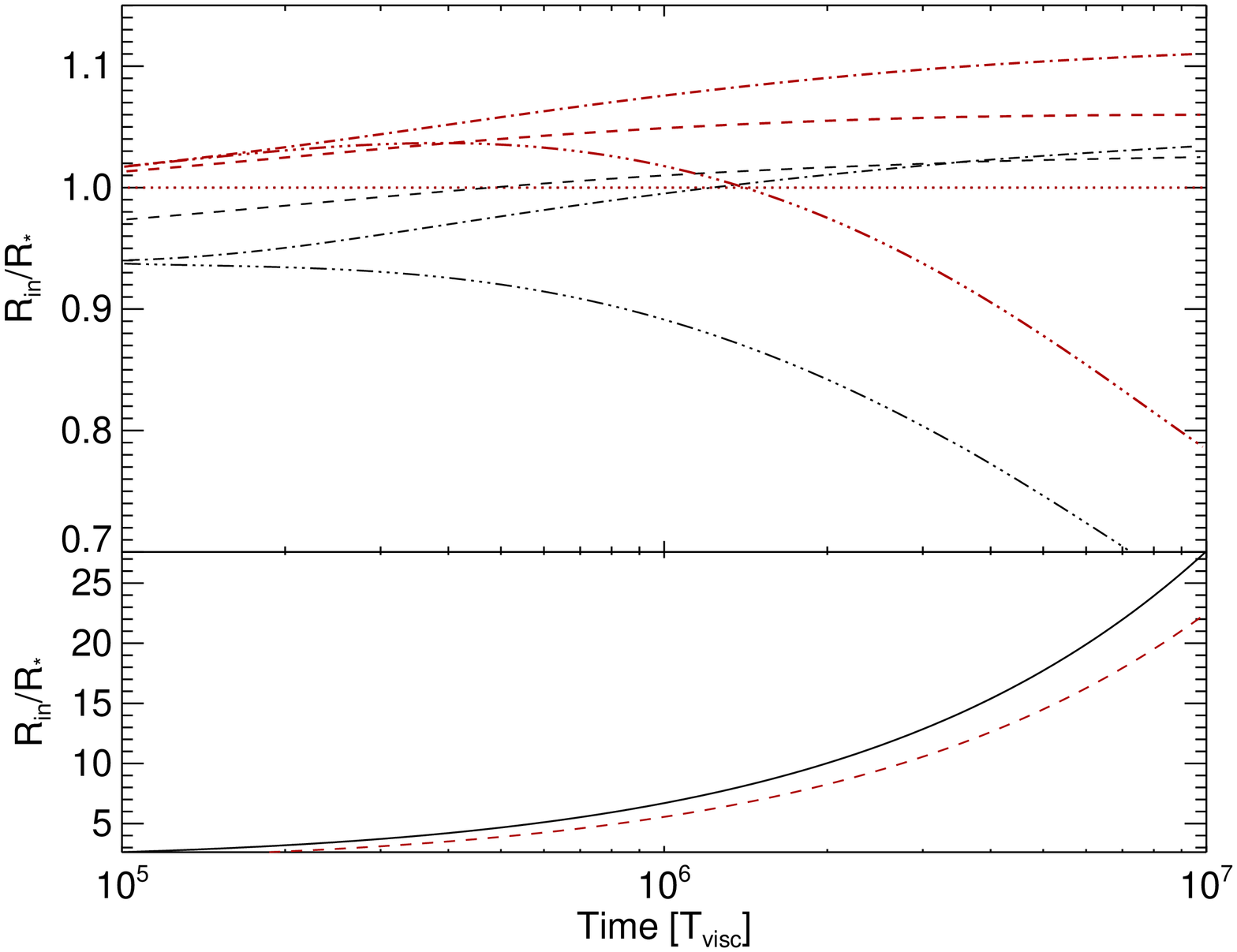}}}
     \caption{Evolution of \rin\ and \rc\ in a dead disc with
        different \rout. Bottom: The evolution of \rin\ (solid black
        curve) and \rc\ (dashed red curve) for the smallest disc,
        $r_{\rm out,0} = 10~r_{\rm in,0}$. Top: The evolution of \rin\
        (thin black curves) and \rc\ (thick red curves) for different
        sizes of disc, divided by the $r_{\rm out,0} = 10~r_{\rm
          in,0}$ solution. The individual curves are the same as in
        Fig. \ref{fig3:const_ang1}.\label{fig3:const_ang2}}
\end{figure}

\bhl We adopt\ehl\ the representative model parameters used before, with an initial 
inner radius set to $1.3 r_{\rm c}$ (corresponding to a negligible accretion
rate for our chosen $\Delta r_2/r_{\rm in}$), and set $r_{\rm out,0} =
[10,100,10^3,10^4]\,r_{\rm in,0}$. The evolution of \rin\ and \rc\ is
plotted in Figs.~\ref{fig3:const_ang1} and
\ref{fig3:const_ang2}. Fig. \ref{fig3:const_ang1} compares the evolution
of $r_{\rm in}/r_{\rm c}$ for different sizes of disc, while
fig. \ref{fig3:const_ang2} compares the evolution of \rin\ and \rc\ for
different disc sizes to the simulation where $r_{\rm out,0} = 10r_{\rm in,0}$. In
both simulations, the different curves correspond to different initial \rout:
10 \rin (dotted curve), 100 \rin\ (dashed curve), 1000 \rin\
(dash-dotted curve), $10^4 r_{\rm in,0}$ (dash-triple-dotted curve).

Fig.~\ref{fig3:const_ang1} shows the evolution of $r_{\rm in}/r_{\rm  c}$ in 
time for different initial \rout. For the simulations with $r_{\rm out,0}/r_{\rm in,0} = 
[10,100,1000]$, the ratio $r_{\rm in}/r_{\rm c}$ declines to a minimum 
value that decreases as \rout\ \bhl is taken\ehl\ larger, before again 
increasing \bhl approximately\ehl\ logarithmically. In the largest disc, 
$r_{\rm out,0}/r_{\rm in,0} = 10^4$, the evolution is the same as in the smaller
discs at early times, but $r_{\rm in}/r_{\rm c}$ continues to decline
for much longer until it reaches a minimum at around $10^7t_*$ when it
finally turns over.

The minimum value of $r_{\rm in}/r_{\rm c}$ is thus determined by the
amount of mass in the disc available for accretion. The larger discs 
have more mass, which sustains the accretion rate onto the star for a 
longer time before the drop in surface density causes \rin\ to move 
outward again. 

The bottom panel of Fig.~\ref{fig3:const_ang2} shows the evolution of
\rin\ and \rc\ for $r_{\rm out,0} = 10~r_{\rm in,0}$. The evolution is
qualitatively the same as we derived in the analytic approximation in
Sec. \ref{sec3:ansol}. Initially \rin\ remains fixed as \rc\ starts to
evolve outwards, until \rc\ moves close enough to \rin\ that accretion
can begin. The inner radius then evolves outward at approximately the
same rate as \rc\ as the star spins down. The variation between
different simulations is emphasized in the top panel of
Fig.~\ref{fig3:const_ang2}.  Here we plot the evolution of \rc\ (thick
curves, red) and \rin\ (thin curves, black) for the discs with
$r_{\rm out,0}/r_{\rm in,0} = [10^2, 10^3, 10^4]$, divided by the solution for
$r_{\rm out,0}/r_{\rm in,0} = 10$. For larger discs the accretion rate is
higher, so that \rin\ can move closer to \rc. In the three smaller
discs, the accreted mass adds a negligible amount of angular momentum
to the star, so that as \rin\ moves closer to \rc. As a result, the
spin-down torque simply becomes more efficient, and \rc\ spins down
faster. After $10^7 t_*$ \rc\ for the disc with $r_{\rm out,0}/r_{\rm
  in,0} = 10^3$ is more than 10\% larger than in the smallest
disc. However, for the largest disc, the accretion rate puts \rin\
close enough to \rc\ that the spin-down torque starts to drop in
efficiency and the spin-up from accretion becomes
non-negligible. Although \rc\ still increases, after $10^7 t_*$ \rc\
is 30\% smaller than for a small disc.

These results show that size of the disc can considerably influence
the efficiency of spin-down, emphasizing the fact that the spin-down
of a star is an initial value problem. The initial size of the disc
can be as important as the ratio $T_{\rm SD}/T_{\rm visc}$ and the
parameter $\Delta r_2$ in determining whether the disc can become
trapped, and the efficiency of the spin-down torque. The results of
this section would suggest that larger discs (with their larger
reservoirs of mass) are more likely to become trapped than smaller
discs. 

\begin{figure}
\rotatebox{90}{\resizebox{!}{84mm}{\includegraphics[width=84mm]{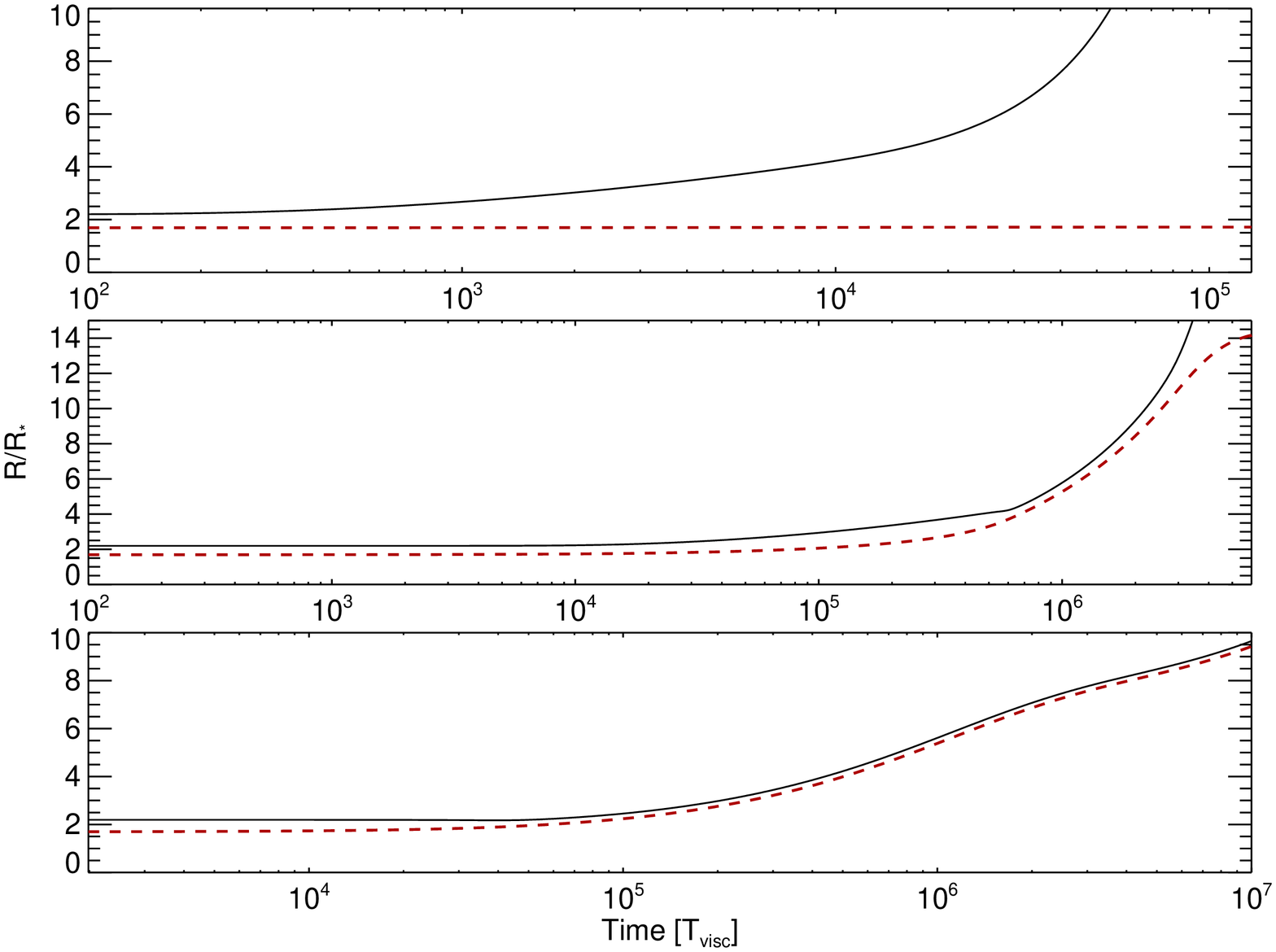}}}
     \caption{Top: Evolution of \rin\ and \rc\ in a dead disc with mass transport
        through \rout, where $T_{\rm visc}(r_{\rm out,0}) \gg T_{\rm SD}$. The
        moment of inertia in the disc is sufficiently large to prevent
        \rin\ from moving out as \rc\ evolves. Middle: Evolution of \rin\ and \rc\ in a
        dead disc with mass transport through \rout, where $T_{\rm
          visc}(r_{\rm out,0}) \sim T_{\rm SD}$. The disc is initially massive
        enough to spin down the star, but after some time evolves away from
        the equilibrium solution and \rin\ moves rapidly out. Bottom: Evolution of
        \rin\ and \rc\ in a dead disc with mass transport through \rout, where
        $T_{\rm visc}(r_{\rm out,0}) \gg T_{\rm SD}$. The moment of inertia in
        the disc is sufficiently large to prevent \rin\ from moving out as
        \rc\ evolves. Curves show \rin\ and \rc\ as in fig. \ref{fig3:stable_visc}.\label{fig3:sig0}}
\end{figure}

\subsection{Long-term behaviour of discs of finite size}
\label{sec3:disc_spread}

\bhl As we found in section \ref{sec3:trapuntrap}, the long-term evolution 
of the disc+star system tends to `bifurcate'. The end state is either a
star that has lost little angular momentum, surrounded by a dead disc, 
or a star continuously spun down by a disc trapped at corotation with 
the star.

We investigate this further by a set of simulations in which the
initial state is a disc of finite size $r_0$, evolving in a grid that
is 10 times larger, \rout$=10\, r_0$ \ehl. Apart from this change we
use the standard parameters (section \ref{sec3:standard})\ehl, setting
the \bhl initial value of $r_{\rm in}/ r_{\rm c}$ at 1.3\ehl\ (for
comparison with the results of the previous section). Whether the disc
will be able to substantially spin down the star depends on the ratio
the timescales $T_{\rm visc}$ at $r_0$ to $T_{\rm SD}$. If the disc is
very small ( $T_{\rm visc}(r_0) \ll T_{\rm SD}$), \rin\ will move
outward too quickly to spin down the star, while if $T_{\rm visc}(r_0)
\gg T_{\rm SD}$, the moment of inertia in the disc is sufficiently
high to be able to absorb much more angular momentum from the star, so
that the disc can operate as an efficient sink for the star's angular
momentum.

Fig.~\ref{fig3:sig0} shows results for three different sizes of
disc. In each curve, the dashed red line shows the evolution of \rc,
while the solid black curve shows the evolution of \rin.  The top
panel of fig. \ref{fig3:sig0} shows the disc evolution when the
viscosity is chosen such that $T_{\rm visc}(r_0) \ll T_{\rm SD}$, with
$r_0 = 100\,r_{\rm in}$. Since the angular momentum injected at \rin\
is carried away by viscous spreading of the disc, the disc quickly
evolves away from its initial configuration. Since the moment of
inertia in the disc is much smaller than in the star, the disc becomes
too spread out to absorb the angular momentum injected at \rin, and
\rin\ moves outward before the star is able to slow down
substantially.

In the middle panel of \ref{fig3:sig0} we show the evolution of
\rin\ and \rc\ when the spin-down timescale is comparable to the viscous
timescale at $r_0$. The disc also initially diffuses outwards (as seen
by the increase in \rin\ around $t = 5\times10^5$), so that the rate of
angular momentum exchange from the star to \rin\ decreases. This allows
\rc\ to catch up, and the two radii start to evolve
together. Eventually, however, viscous spreading of the disc wins, and
\rin\ begins to move outwards.

The bottom panel of \ref{fig3:sig0} shows the evolution of \rin\ and
\rc\  when $T_{\rm visc}(r_0) \gg T_{\rm SD}$. The result is
essentially the same as in Sec. \ref{sec3:const_am}, since the disc is now
so large that additional angular momentum from the star \bhl does not
reach \rout\ on the spindown timescale\ehl. In other words, the
moment of inertia of the disc itself is large enough that the star is
able to spin down without causing \rin\ to move rapidly outward.

\subsubsection{The rotation of Ap stars and magnetic white dwarfs}
\label{sec3:Apstars}

An intriguing clue to the spindown of magnetic stars comes
from slowly rotating Ap stars. As a class these stars are observed to
have very strong dipolar magnetic fields (up to 10 kG). \bhl A few of 
them have extremely long rotation periods (up to 10-100 years),
while others have rotation periods as short as 0.5 d\ehl. A similar 
phenomenon is observed in the magnetic white dwarfs. Most
of these have periods of a few days or weeks, but some rotate as 
slowly as the slowest Ap stars. We suggest that the bifurcation of 
outcomes we found in the above is the underlying reason 
for the remarkable range of spin periods of magnetic stars.

\section{Conclusions}

\bhl As found before in \cite{1977PAZh....3..262S} and DS10, a disc in
contact with the magnetosphere of a star can be in a `dead' state,
with its inner edge well outside the corotation radius so the
accretion rate onto the star vanishes, and the torque exerted by the
magnetic field transmitted outward by viscous stress. In the
calculations reported in DS10, an additional state was found,
intermediate between the accreting and dead state. In this state, the
inner edge of the disc stayed close to the corotation radius \rc, even
as the accretion rate onto the star varied by large factors. We call
this phenomenon {\em trapping} of the disc. Accretion can be
stationary or in the form of a limit cycle in this state.

One of the goals of this investigation was to find out under what
conditions this trapping takes place, using a series of numerical
experiments with varying initial and boundary conditions. If initial
conditions are such that the inner edge starts inside corotation and
slowly moves outward, we find that the disc gets stuck in a trapped
state for a long time if the disc viscosity $\nu$ is low (up to about
$10^3$ times shorter than the spindown timescale of the star). The
accretion rate then slowly vanishes but the inner edge always stays
close to corotation. At higher viscosity ($> 10^5\times$ the spindown
timescale) on the other hand, the disc evolves through corotation into
a `dead' disc state, with inner edge well outside the corotation
radius. In terms of the standard viscosity parametrization $\nu=\alpha
(H/r)^2$, compared with the spin-down timescale of the disc, a trapped
state is more likely to happen in strongly magnetic ($B_{\rm S} \sim
10^{12}G$) X-ray pulsars and protostellar discs than the weaker
magnetic fields ($B_{\rm S} \sim 10^{8}G$) of millisecond X-ray pulsars.

Our second goal was to find out how a star spins down in the long term, 
under the influence of the angular momentum it loses by the magnetic 
torque exerted on the disc. The results show an interesting `bifurcation'
of long-term outcomes: if the disc evolves into a dead state, the star loses 
only a fraction of its initial angular momentum, and can remain spinning
rapidly throughout its life. If on the other hand it enters a trapped state  
at some point, it  remains in this state. The star can then slow down to very 
low rotation rates, the inner edge of the disc tracking the corotation radius 
outward. We suggest that these two outcomes can be identified
respectively with the rapidly rotating and slowly rotating classes of magnetic 
Ap stars and magnetic white dwarfs. The evolution of the trapped
state could also be reproduced with a simplified model that does not require
solving the full viscous diffusion equation.

This picture of magnetosphere-disc interaction differs from the
standard view that mass will be `propellered' out of the system
instead of accreting, once the star rotates more rapidly than the
inner edge of the disc. As shown already by \cite{1977PAZh....3..262S}
and again argued in \cite{1993ApJ...402..593S} and DS10 this
assumption is not necessary, in many cases unlikely, and ignores some
of the theoretically and observationally most interesting aspects of
the disc-magnetosphere interaction\ehl.

Though \bhl the dead state\ehl is a regime where a significant fraction of the disc mass 
could in principle be expelled (\rin $\gg$ \rc), the results presented here show 
that \ehl\ magnetospherically accreting systems \bhl often avoid\ehl\ this regime.
Instead, they end up in the trapped state, in which the disc-field
interaction keeps the inner radius truncated very close to the
co-rotation radius, even at very low accretion rates. \bhl Both the trapped
and (\rin $\approx$ \rc) and the dead state (\rin $\gg$ \rc) allow the
disc to efficiently spin down the star. The disc retains a large amount of
mass, but in the absence of accretion onto the central star appears 
quiescent.\ehl

\section{Discussion}

\bhl
By assuming a given dependence of viscosity on distance, we have
bypassed the physics that determines it. In terms of the standard
 $\alpha$-parametrization of viscosity, we have left out the physics
 that determines the disc temperature and hence its thickness $H/r$.
 Additional time dependence or instabilities may arise from feedback
 between accretion and disc temperature. The radiation produced by 
 matter accreting on the star could be large enough to change temperature, 
 the ionization state and hence the thickness at the inner edge of the disc. 
 In a trapped disc, this is just the region that controls the accretion rate 
 onto the star. The size of the transition region, 
 which we have parametrized with the widths $\Delta r, \Delta r_2$ may 
 well depend on disc thickness. Positive feedback may be thus possible.
 We leave this possibility for future work.\ehl

\bhl In systems such as the accreting X-ray pulsars the accretion is
episodic on long timescales. This is attributed to the instability of viscous 
discs that is also responsible for the outbursts of cataclysmic variables. 
In these cases, in\ehl\ which there is a large change in $\dot{m}$ and 
\rin\ is far from \rc\ in the quiescent phase, 
there is the possibility of hysteresis: the same accretion rate will lead 
to a different value of \rin\ (and therefore disc torque) depending on 
whether the source is moving into or out of outburst. As the source 
goes into outburst, the disc will not have as much mass in its inner 
regions, so that \rin\ will move inward gradually from large radii until it
crosses \rc\ and the outburst begins. In the decline phase, however,
the disc will become trapped around the inner radius of the disc when
the accretion rate drops, allowing for a larger spin-down in the disc
and accretion bursts via the instability of DS10. \bhl  The net effect of  
such episodic accretion on the spin history of the star, as compared
with the case of steady accretion, is not obvious\ehl. 
We discuss this in more depth in the companion paper.


\bhl Some work on disc-magnetosphere interaction assumes magnetic
torques to act over a significant part of the disc\ehl \citep{1991ApJ...370L..39K,
1996MNRAS.280..458A}. More recent work (and in particular numerical
MHD simulations of the disc-field interaction) finds the interaction region 
to be much narrower \bhl(as we have also assumed here), and spindown
torques on the star are correspondingly smaller. The difference for the
long-term spin evolution is not dramatic, however, as our trapped disc
results demonstrate.\ehl


Interestingly, \cite{1996MNRAS.280..458A} also observed that their
discs would become trapped around \rc\ as the accretion rate in the
disc decreased by several orders of magnitude. In their model, the
magnetic field-disc interaction also acts like a boundary condition at
low accretion rates and the disc evolved viscously in response. This
is presumably because although the disc in their model is threaded by
a magnetic field everywhere, the disc-field interaction is by far the
strongest in the inner regions, so that they see a similar behaviour
to the one described in this paper.

\bhl MHD simulations of interaction between the disc and the magnetic 
field are becoming increasingly realistic. 
These simulations can only run for\ehl\ very short timescales (the longest
 of order $T_{\rm visc}$ at \rin), so they tend to emphasize initial
transients. Still, they offer insight in how the disc and magnetic field will 
interact. To date, most simulations have concentrated on strongly accreting
\citep{1996ApJ...468L..37H,1997ApJ...489..199G,1997ApJ...489..890M} or
propellering \citep{2004ApJ...616L.151R} cases.
\bhl Simulations that come closest to the conditions of a trapped 
disc are\ehl\ the study of a so-called `weak propeller' regime by
\cite{2006ApJ...646..304U}. The authors found that discs
in which \rin\ was initially truncated close to \rc\ launch much
weaker outflows than discs truncated further away. They also found
some evidence of the field changing the disc structure (as shown from
mass piling up in the inner regions), although in their simulation the
majority of the angular momentum in the star was carried away via a
wind, rather than through the disc. These simulations also did not run for
very long, however, \bhl so it is hard to separate transient behaviour due 
to the initial conditions from the longer-term systematic effects we 
are interested in.\ehl

\cite{2006ApJ...639..363P} did calculations were done for a case where
the star's field is inclined with respect to the disc axis\bhl. The
authors' results suggest that the transition width which we have
parametrized by $\Delta r_2$\ehl\ would increase with the inclination
of the magnetic field, since there will be some values for \rin\ at
which which both accretion and disc mass trapping can occur. This then
would suggest that dead discs are more likely to form in systems with
small inclinations between the spin axis and magnetic axis, and also
that the disc instability studied in DS10 (which also tends to occur
for smaller values of $\Delta r_2$). This prediction is supported by
the observation that Ap stars with the longest periods tend to have
the lowest inclination angles for the magnetic field
\citep{2000A&A...359..213L}. \bhl

\bhl The transition widths of disc-magnetosphere interaction that can be
inferred from these simulations are significant, and are in the range we have
assumed here. They are much larger than the very narrow interaction
regions assumed by Matt et al. \citep{2004ApJ...607L..43M, 
2005ApJ...632L.135M, 2010ApJ...714..989M}\ehl.

We have  found that the distance from \rc\ at which the inner edge
of the disc gets trapped is determined by the ratio of two timescales
for the disc's evolution: the viscous evolution timescale in the disc
(which determines the rate at which disc density profile can change)
and the star's spin-down timescale (which sets the rate at which
\rc\ moves outwards). The viscous timescale is in general much shorter
than the spin-down timescale, and the ratio of the two timescales
varies from $\sim 10^{-3}$ (in protostellar discs) to $\sim 10^{-17}$
(in millisecond X-ray pulsars). The larger the ratio, the longer the
disc will take to respond to changes in \rc. As a result \rin\ remains
closer to \rc\ than it would if the ratio were smaller, and the trapped
disc has a higher accretion rate onto the star. If the ratio is too
low, then \rin\ moves much further away from \rc, and the system
likely enters the \bhl dead disc regime\ehl.


\bhl For the parameters characterizing the interaction region between disc and 
magnetosphere in our model, we have used here values such that the cyclic 
accretion behavior found in DS10 does not develop. This was done
for convenience, since the short time steps needed to follow these cycles 
makes it harder to calculate the long-term evolution.  The 
time-averaged effect of these cycles is not expected to make a big difference
for the long-term evolution. 

These cyclic\ehl\ accretion bursts can 
persist in the trapped state, when the star is spinning down
efficiently. They are observed to occur both over
several orders of magnitude of accretion rates and transiently (over a
small range of accretion rate), and \bhl depend on\ehl\ the ratio
of timescales discussed above. Whether or not the instability occurs
is determined by the detailed disc-field interaction (the parameters
$\Delta r$ and $\Delta r_2$ in our model). The peak of the accretion
bursts is typically much larger ($> 10\times$) than the mean accretion
rate for the system, and the period is typically between $0.01-10^2
T_{\rm visc}(r_{\rm in})$. \bhl The properties and conditions for 
occurrence of these cycles are studied further in a companion paper\ehl.


\subsection{`Propellering'}

\bhl In our calculations we have left out the possibility that
interaction of the magnetosphere with the disc will cause of mass
ejection from the system.  The point being that, contrary to common
belief, such interaction can function without mass ejection by
`propellering', as pointed out already by \cite{1977PAZh....3..262S}.
Understanding of this restricted case, as we have developed here, is
prerequisite for understanding the much less well defined case of mass
loosing discs.

On energetic grounds, mass loss from the system is necessarily
limited, unless the inner edge is well outside corotation
\citep{1993ApJ...402..593S}. This point has also been made by
\cite{2006ApJ...639..363P}, who propose that mass lifted at \rin\ may
fall back on the disc at some finite distance. This would create a
feedback loop in the mass flux through the disc, opening the
possibility of additional forms of time-dependent behaviour that do
not exist in accretion onto non-magnetic stars. In the trapped disc
state we have studied here the difference in rotation between disc and
star is small, so any significant amount of mass kicked up from the
interaction region cannot move very far before returning to the
disc. Its effects are then secondary, at least for the long-term
evolution of the disc.

The possibility of significant effects of mass loss is more realistic for
the dead disc states, where the distance of the inner edge from corotation
can become much larger.

Real propellering is expected to happen when mass transfer from a
companion star sets in for the first time onto a rapidly spinning magnetic 
star. The cataclysmic binary AE Aqr is evidently such a case
\citep{2003MNRAS.338.1067P}. A disc is absent in this CV, and all mass 
transfered appears to be ejected in a complex outflow associated with 
strong radio emission. 
\ehl


\section{Acknowledgements}
CD'A acknowledges support from the National Science and Engineering
Research Council of Canada.
\bibliographystyle{mn2e}
\bibliography{magbib}
\label{lastpage}
\end{document}